\documentclass[useAMS,usenatbib,onecolumn]{mn2e}
\usepackage{graphicx,subfigure,enumerate,pstricks,latexsym}
\usepackage{lipsum}
\usepackage[english]{babel}

\def\beq{\begin{equation}} \def\eeq{\end{equation}}
\def\bea{\begin{eqnarray}} \def\eea{\end{eqnarray}}

\newcommand{\Schw}{Schwarzschild}

    \def\mir{\mathrm{r}} \def\mit{\mathrm{\theta}}

\def\fl{}

\title[Limits on mass of the intermediate black hole]{Mass of intermediate black hole in the source M82 X-1 restricted by models of twin high-frequency quasiperiodic oscillations}

\author[Z. Stuchl\'{\i}k and M. Kolo\v{s}]
{Z.~Stuchl\'{\i}k$^{1}$\thanks{zdenek.stuchlik@fpf.slu.cz} and M.~Kolo\v{s}$^{1}$\thanks{martin.kolos@fpf.slu.cz}\\
$^{1}$Institute of Physics, Faculty of Philosophy \& Science, Silesian University in Opava,\\ Bezru\v{c}ovo n\'{a}m\v{e}st\'{i} 13, CZ-74601 Opava, Czech Republic}

\begin{document}



\maketitle

\label{firstpage}

\begin{abstract}
We apply the relativistic precession model with its variants and the resonance epicyclic model with its variants, based on the frequencies of the geodesic epicyclic motion in the field of a Kerr black hole, to put limits on the mass of the black hole in the ultraluminous X-ray source M82 X-1 demonstrating twin high-frequency quasiperiodic oscillations (HF QPOs) with the frequency ratio near 3:2. The mass limits implied by the geodesic HF QPO models are compared to those obtained due to the model of string loop oscillations around a stable equilibrium position. Assuming whole the range of the black hole dimensionless spin, $0<a<1$, the restrictions on the black hole mass related to the twin HF QPOs are widely extended and strongly model dependent, nevertheless, they give the lower limit $M_{\rm M82X-1}>130~M_{\odot}$ confirming existence of an intermediate black hole in the M82 X-1 source. The upper limit given by one of the variants of the geodesic twin HF QPO models goes up to $M_{\rm M82X-1}<1500~M_{\odot}$. The range 37~mHz-210~mHz of the low frequency QPOs observed in the M82~X-1 source introduces additional restrictive limits on the black hole mass, if we model the low frequency QPOs by nodal precession of the epicyclic motion. The nodal precession model restrictions combined with those implied by the geodesic models of the twin HF QPOs give allowed ranges of the M82 X-1 black hole parameters, namely $140~M_{\odot}<M_{\rm M82X-1}<660~M_{\odot}$ for the mass parameter, and $0.05<a_{\rm M82X-1}<0.6$ for the spin parameter.
\end{abstract}


\section{Introduction}\label{intro}

Importance of finding a $3:2$ twin HF QPOs in an ultraluminous X-ray source in order to resolve the issue of an intermediate mass black hole has been pointed out by \cite{Abr-etal:2004:ApJ:}. Quite recently, such stable twin-peak X-ray quasi-periodic oscillations with 3:2 frequency ratio have been reported for the ultraluminoues M82~X-1 source; the observed frequencies are large enough to be interpreted as twin HF QPOs if the assumed central black hole in the source has an intermediate ($M>100~M_{\odot}$) mass \citep{Pas-Str-Mus:2015:Nat:}. Estimates of the M82~X-1 mass were made by \cite{Pas-Str-Mus:2015:Nat:} being based on the simple mass scaling method or on a Monte Carlo approach related to the relativistic precession model \citep{Ste-Vie:1999:PHYSRL:} that has been applied for fitting the timing measurements data obtained in the microquasar GRO~1655-40 \citep{Mot-etal:2014:MNRAS:} where also related low frequency QPOs have been taken into account being modelled by the nodal precession \citep{Ste-Vie:1999:PHYSRL:}. While in the GRO~1655-40 microquasar one of the low frequency QPOs is clearly related to the simultaneously observed twin HF QPO, in the case of M82~X-1 ultraluminous source, we cannot uniquely identify one of the observed low frequency QPOs that could be directly related to the twin HF QPOs. For this reason, the estimates of the M82~X-1 black hole mass cannot be precise if based only on the timing measurements, as we can only assume a common radius of the measured twin HF QPOs and a low frequency QPO being related to the whole set of the observed low frequency QPOs. 

We consider useful and necessary to realize a more extended and robust analysis of the limits on the M82~X-1 intermediate black hole mass using a variety of HF QPO models based on the frequencies of the geodesic (quasi)circular motion, i.e., the Keplerian frequency, or the radial and vertical epicyclic frequencies of the test particle motion in the Kerr geometry. We select the geodesic models that are able to explain the twin HF QPOs observed with the $3:2$ frequency ratio in the microquasars (binary systems containing a black hole) GRS 1915+105, GRO 1655-40, XTE 1550-564, assuming that these models could be relevant also for the scaled up intermediate mass black hole. \footnote{It should be stressed that none of the geodesic models is capable to explain the $3:2$ twin HF QPOs in all of the three microquasars \citep{Tor-etal:2011:ASTRA:}.} Along with the successful geodesic models we apply also some of their variants that are not able to explain the $3:2$ twin HF QPOs in the three microquasars, but could be interesting for the ultraluminous source M82 X-1. 

To obtain the limits on the black hole mass, we use the resonance relations technique introduced in \citep{Stu-Kot-Tor:2013:ASTRA:,Stu-Kot-Tor:2011:ASTRA:}. For each of the geodesic oscillation models this technique introduces a specific dependence of the dimensionless radius $x$ where the $3:2$ twin HF QPOs occur on the dimensionless black hole spin $a$. This $x(a)$ dependence is governed by the frequency ratio, being independent of the black hole mass $M$ that is controlled by magnitude of the observed frequencies as shown in \citep{Stu-Kot-Tor:2013:ASTRA:}. We also assume that some of the observed low frequency QPOs can be related to the same radius as the twin HF QPOs, applying thus the resonance relations technique to the nodal (Lense-Thirring) frequency model along with the considered variant of the geodesic model of twin HF QPOs. The low frequency QPOs can be then used to put additional restrictions on the black hole mass and the black hole spin, if combined with the restrictions implied by the geodesic twin HF QPO models. 

We compare the M82 X-1 mass limits obtained in the framework of the HF QPO models based on the frequencies of the geodesic, test particle motion that are determined purely by the gravitational field of the black hole to the mass limits obtained in the framework of a non-geodesic model, namely the string loop oscillation model \citep{Stu-Kol:2014:PHYSR4:}. The axisymmetric current-carrying string loops could represent plasma exhibiting a string-like behaviour due to dynamics of the magnetic field lines \citep{Sem-Dya-Pun:2004:Sci:,Chri-Hin:1999:PhRvD:}, or due to the thin flux tubes of magnetized plasma simply described as 1D strings \citep{Sem-Ber:1990:ASS:,Cre-Stu:2013:PhRvE:,Cre-Stu-Tes:2013:PlasmaPhys:,Kov:2013:EPJP:}. Their oscillations are thus governed not only by the black hole gravitational field, but also the string loop parameters determining the interplay of the string loop angular momentum and (magnetic) tension are relevant \citep{Stu-Kol:2012:JCAP:,Kol-Stu:2013:PHYSR4:}. Contrary to the geodesic models, the string loop oscillation model can explain the $3:2$ twin HF QPOs in all the three microquasars \citep{Stu-Kol:2014:PHYSR4:}, moreover, it can be applied also to neutron star systems \citep{Stu-Kol:2015:GRG:}. It is thus interesting to test its predictions for the intermediate black hole system.

\section{Twin HF QPOs observed in the ultraluminoues M82 X-1 source}

M82 X-1 is the brightest X-ray source in M82 galaxy, giving extremely high luminosity indicating a central black hole with intermediate mass in the interval $10^2-10^4~M_{\odot}$ \citep{Mat-etal:2001:ApJ:,Kaa-etal:2001:MNRAS:,Dew-Tit-Gri:2006:ApJ:,Muc-etal:2006:MNRAS:,Cas-etal:2008:MNRAS:}. The mass estimates inferred from the luminosity measurements were later confirmed by estimates based on modelling the continuum X-ray spectrum of the source that put the mass limits in the interval $200-800~M_{\odot}$ \citep{Fen-Kaa:2010:ApJ:}; however, there is a model predicting a substantially lower mass $\sim20~M_{\odot}$ \citep{Oka-Epi-Kaw:2006:ApJ:}.

Another information on the mass of the M82 X-1 black hole is contained in the important observations of low-frequency quasiperiodic oscillations \citep{Str-Mus:2003:ApJ:}. For sources containing a black hole, the low-frequency QPOs are usually related to the nodal (Lense-Thirring) frequency of the test particle motion \citep{Ste-Vie:1999:PHYSRL:}. \footnote{Recently a model of the low-frequency quasiperiodic oscillations based on the 'rocking frequency' of toroidal discs is also discussed \citep{Axe-Don-Hja:2008:MNRAS:}.} However, the mass estimates based on scaling relations using low-frequency characteristic time scales, are strongly uncertain \citep{Pas-Str:2013:ApJ:,Pas-Str-Mus:2015:Nat:}. Therefore, the observational results related to the stable twin HF QPOs reported quite recently in \citep{Pas-Str-Mus:2015:Nat:} are extremely important as they bring a possibility of obtaining more stringent restrictions on the mass of the central black hole as predicted by \cite{Abr-etal:2004:ApJ:}, giving potentially a clear information on existence of an intermediate mass black hole. Really, magnitude of the observed frequencies and character of the oscillations indicate strongly that they should be twin HF QPOs occurring in close vicinity of the black hole horizon, being related to the orbital motion. The twin HF QPOs observed in the M82 X-1 source are similar to those observed in the microquasars GRS 1915+105, GRO 1655-40, XTE 1550-564. There is the same stable $3:2$ frequency ratio of the observed frequencies, and magnitude of the frequencies is comparable to the magnitude of the orbital (Keplerian) frequency of a test particle orbiting at the innermost stable circular geodesic, if we assume an intermediate black hole at the M82 X-1 source. Moreover, the twin HF QPOs could be related also to the low-frequency quasi-periodic oscillations observed in the M82 X-1 source \citep{Pas-Str-Mus:2015:Nat:}. 

The lower and the upper frequency of the twin quasiperiodic oscillations observed at the source M82 X-1 as presented in \cite{Pas-Str-Mus:2015:Nat:} reads 
\beq
 f_{\rm L} = 3.32\pm0.06~\mathrm{Hz}, \quad f_{\rm U} = 5.07\pm0.06~\mathrm{Hz}. \label{ffL} \label{ffU}
\eeq
Because of the mass-scaling of the quasiperiodic oscillations observed in the microquasars, these frequencies should correspond to the HF QPOs in the field of intermediate mass black holes \citep{Tor-etal:2005:ASTRA:}. The other details of the detected twin HF QPOs, namely the fractional rms amplitudes and the quality factors, are presented in \citep{Pas-Str-Mus:2015:Nat:}. We do not discuss them here, since we concentrate our attention on the observed frequencies only. 

The low frequency oscillations observed in the source M82 X-1 are also fully taken into account in \citep{Pas-Str-Mus:2015:Nat:}. The frequency range and mean value of this range 
\beq
 37~\mathrm{mHz} < f_{\rm low} < 210~\mathrm{mHz}, 
\quad f_{\rm low-mean} \sim 120~\mathrm{mHz}, \label{fflow} \label{fflowM}
\eeq
has been also considered in the estimates of the intermediate black hole mass presented in \citep{Pas-Str-Mus:2015:Nat:} where a Monte Carlo technique has been used that was developed for application of the relativistic precession model to explain the twin HF QPOs and simultaneously observed low frequency QPO in the microquasar GRO~J1655-40 \citep{Mot-etal:2014:MNRAS:}. The M82~X-1 black hole mass estimated in this way reads 
\beq
 M_{\rm M82X-1} \sim 415\pm63~M_{\odot}. 
\eeq
However, contrary to the case of the GRO~J1655-40 microquasar \citep{Mot-etal:2014:MNRAS:}, for the low frequency QPOs considered in \citep{Pas-Str-Mus:2015:Nat:} we are not able to fix the frequency of the low-frequency QPOs simultaneously observed with the twin HF QPOs. Therefore, the error in establishing the M82 X-1 black hole mass is large as all the observed low frequency QPOs are considered as a possible partner of the twin HF QPOs. 

\section{Models of twin HF QPOs}

We shall first give a short overview of the models of twin HF QPOs in low-mass X-ray binary (LMXB) systems. The twin HF QPOs are observed in both neutron star and black hole LMXBs, but we concentrate on the black hole cases (microquasars) that could be extended to the systems with an intermediate or a supermassive black hole. In such systems we can always assume a crucial role of the strong gravity in vicinity of the black hole horizon as the observed frequencies are close to the orbital and epicyclic frequencies of the quasicircular motion of test particles in the field of black hole with mass determined by methods independent of the timing measurements. \footnote{Most of the discussed models can be also applied to the neutron star LMXBs, although they require modified approach in comparison to the black hole systems because of the presence of the neutron star surface \citep{Stu-Kot-Tor:2013:ASTRA:}.}

The models of twin HF QPO in the black hole systems can be separated into three categories. 

\subsection{Hot spot kinematic models} \label{GOmodels}

The kinematics of the epicyclic orbital motion allows for variability related to the motion of "hot spots", i.e. radiating blobs of matter in the innermost parts of the accretion disc. The standard relativistic precession (RP) model \citep{Ste-Vie:1998:ApJ:,Ste-Vie:1999:PHYSRL:} demonstrates the two modes of the relativistic epicyclic motion of hot spots as the twin HF QPOs. The two modes are related to the orbital motion and periastron (radial) precession of the relativistic orbits in strong gravity near the black hole horizon. Moreover, due to the Lense-Thirring relativistic precession, the RP model can be related also to the low-frequency QPO modes observed on time scales about one order of magnitude slower than the twin HF QPOs \citep{Ste-Vie:1998:ApJ:}. Considering the epicyclic motion in both radial and vertical direction, modifications of the RP model can be defined, e.g., the so called total precession model \citep{Stu-Kot-Tor:2013:ASTRA:}. 

An alternative to the RP model is the tidal disruption (TD) model \citep{Cad-Cal-Kos:2008:ASTRA:,Kos-etal:2009:ASTRA:} according to which the twin HF QPOs are generated by large accreting inhomogeneities deformed to an orbiting "ring-section" by tidal forces of the black hole. 

Note that in the RP and TD models no explanation of the observed $3:2$ frequency ratio exists at the present state of knowledge, however, one cannot exclude existence of some resonant phenomena between the two modes of oscillations. We can only state that in the RP model the $3:2$ frequency ratio is obtained at radii very close to the innermost stable circular orbit (ISCO) where the edge of the Keplerian disc is assumed, while for the TD model this radius is shifted slightly above the ISCO but it is still the innermost region of the accretion disc \citep{Tor-etal:2011:ASTRA:}. 

\subsection{Resonant models}

The resonant models of the twin HF QPOs assume a particular parametric or non-linear forced resonance of the oscillatory modes of the accretion disc \citep{Ali-Gal:1981:GRG:,Abr-Klu:2001:AA:,Abr-Bul-Bur-Klu:2003:ASTRA:,Stu-Kon-Mil-Hle:2008:ASTRA:,Hor-etal:2009:ASTRA:}. The frequency commensurability is thus crucial ingredient of the resonant models, and a particular case of this commensurability occurs for the parametric (internal) resonant phenomena that become strongest in the case of the $3:2$ frequency ratio \citep{Tor-etal:2005:ASTRA:}. The simplest variant of the resonant model is the resonant epicyclic model where the two modes in resonance are identified to correspond to the radial and vertical epicyclic oscillations \citep{Tor-etal:2005:ASTRA:}. Of course, in the case of the non-linear forced resonances, the combinational (beat) frequencies can be relevant \citep{Lan-Lif:1969:Mech:,Nay-Moo:1979:NonOscilations:}. 

The commensurability of the frequencies is crucial also for the model of warped disc (WD) oscillations where resonances are relevant for oscillations of deformed Keplerian discs and the frequencies of the oscillatory modes are governed by combinations of the Keplerian and epicyclic frequencies \citep{Kat:2004:PASJ:,Kat:2008:PASJ:}. Similarly, oscillations of slender tori can be also determined by the frequencies of the geodesic motion \citep{Rez-etal:2003:MNRAS:,Mon-Zan:2012:MNRAS:,Str-Sra:2009:CLAQG:}. However, for non-slender tori, the non-geodesic influence have to be introduced, modifying thus the simple geodesic formulae for the oscillatory frequencies \citep{Str-Sra:2009:CLAQG:}.

\subsection{Discoseismic models}

In the kinetic hot spot models and the resonant models it is assumed that both the observed oscillations are produced by the same mechanism at the same radius of the accretion disc and correspond to interacting modes. On the other hand, in the discoseismic models considering the non-geodesic pressure influence, the oscillatory modes are assumed to be inertial-gravity, corrugation, or pressure, and they are excited by different mechanism at different radii of the disc, being thus non-interacting modes that evolve independently \citep{Kat-Fuk:1980:PAJS:,Wag:1999:PHYSR:,Wag-etal:2001:ApJ:,Zan-etal:2005:MNRAS:}. 

In the case of the models with non-interacting oscillatory modes the $3:2$ frequency ratio condition can be satisfied only for a specific value of the dimensionless black hole spin $a$ \citep{Tor-etal:2011:ASTRA:}. We shall not consider in the following such models, similarly to the HF QPO models based on the specific character of the near-extreme Kerr spacetimes that can be relevant only for near-extreme values of the black hole spin $a \sim 1$ \citep{Stu-etal:2005:PHYSR4:,Stu-Sla-Tor:2007:ASTRA:}. 

Disc oscillations modelled by hydrodynamic simulations of accretion \citep{Zan-etal:2005:MNRAS:,Rey-Col:2009:ApJ:} are shown to be damped by magnetic fields \citep{Fu-Lai:2009:ApJ:,Fu-Lai:2011:MNRAS:}. Moreover, the complex magnetohydrodynamics simulations of accretion processes do not reproduce any phenomena that could be related to the $3:2$ frequency ratio HF QPOs observed in the LMXBs. It is not clear at the present state of knowledge if the magnetohydrodynamics simulations include all the relevant ingredients -- low frequency QPOs were observed in the magnetohydrodynamics simulations including radiative cooling \citep{Mac-Mat:2008:PAJC:}. On the other hand, the twin HF QPOs quite naturally occur in the framework of the string loop oscillation model where the string tension could represent in an approximative way the influence of internal magnetic field of accreting structures \citep{Stu-Kol:2014:PHYSR4:}. 

\section{Geodesic oscillation models}

We can conclude that there is a large variety of the twin HF QPO models, however, no commonly accepted theory of QPO exist at the present state of knowledge \citep{Kli:2006:Compact:}. Nevertheless, most of the models assume the crucial role of gravity and the frequencies of the oscillatory modes are then directly related to the frequencies of the geodesic epicyclic motion or to their combinations. We shall focus attention to these geodesic oscillation models of twin HF QPOs assuming that the black hole spacetime is determined by the standard Kerr geometry. 

We present now the geodesic oscillation models selected here to put limits on the intermediate mass black hole at the M82 X-1 source. More details on the geodesic oscillation models can be found in \citep{Stu-Kot-Tor:2013:ASTRA:}. For each of the selected models we state, if the model can explain the twin HF QPOs in the three microquasars. The mass and spin of the black hole implied by the twin HF QPOs have to be confronted to the mass limits implied by optical measurements independent of the X-ray timing measurements, and the dimensionless spin measurements based on the X-ray observations of spectral continuum or profiled spectral lines -- for details see \citep{Tor-etal:2011:ASTRA:}.

The hot spot models assume radiating hot spots moving along nearly circular geodesic trajectories. In the case of the standard RP model \citep{Ste-Vie-Mor:1999:ApJ:}, the upper of the twin frequencies is attributed to the orbital Keplerian frequency, $\nu_{\rm U}=\nu_{\rm K}$, while the lower one is attributed to the periastron precession frequency, $\nu_{\rm L}=\nu_{\rm K}-\nu_\mir$ where $\nu_\mir$ is the frequency of the radial epicyclic motion. The low frequency QPOs are then related to the nodal (Lense-Thirring) precession with frequency $\nu_{\rm nod}=\nu_{\rm K}-\nu_\mit$ where $\nu_{\theta}$ is the frequency of the vertical epicyclic motion. The RP model can well explain the $3:2$ twin HF QPOs and the related low-frequency QPO at the microquasar GRO 1655-40 \citep{Mot-etal:2014:MNRAS:}. 

Variants of the relativistic precession (hot spot) model are discussed in \citep{Stu-Kot-Tor:2013:ASTRA:,Stu-Kot-Tor:2012:ACTA:}. Here we select the RP1 model introduced by \cite{Bur:2005:RAG:} where the identification of the upper and lower frequencies is given by $\nu_{\rm U}=\nu_{\rm \theta}$ and $\nu_{\rm L}=\nu_{\rm K}-\nu_\mir$, and the RP2 model, called total precession model \citep{Stu-Kot-Tor:2013:ASTRA:}, where $\nu_{\rm U}=\nu_{\rm K}$ and $\nu_{\rm L}=\nu_{\rm \theta}-\nu_\mir$. The RP1 model can explain well the twin HF QPOs at the microquasar XTE 1550-564 \citep{Tor-etal:2011:ASTRA:}, while the RP2 model can explain the twin HF QPOs at the microquasar GRO 1655-40, giving results similar to those of the standard RP model \citep{Mot-etal:2014:MNRAS:}

We extend our selection of the hot spot models of HF QPOs for the TD model based on the idea of an orbiting hot spot distorted by the tidal forces of the black hole \citep{Kos-etal:2009:ASTRA:} where the upper and lower frequencies are identified by the relations $\nu_{\rm U}=\nu_{\rm K}+\nu_\mir$ and $\nu_{\rm L}=\nu_{\rm K}$. The TD model cannot explain the twin $3:2$ twin HF QPOs in any of the three microquasars. However, it could be interesting to use this model for the case of the intermediate mass black hole testing thus the hole tidal effects on large inhomogeneities in the accretion disc.

The resonance epicyclic (RE) models \citep{Abr-Klu:2001:AA:,Tor-etal:2005:ASTRA:,Ali-Gal:1981:GRG:} assume resonance of oscillation modes of accretion discs orbiting black holes giving the simple identification of the upper and lower frequencies $\nu_{\rm U}=\nu_{\rm \theta}$ and $\nu_{\rm L}=\nu_\mir$. This model explain well the twin $3:2$ HF QPOs in the microquasar GRS 1915+105 \citep{Tor-etal:2011:ASTRA:}. The radial profiles of the upper and lower frequencies defined in the framework of the RE and RP models are compared in Figure 1, demonstrating that the $3:2$ frequency ratio occurs closer to the ISCO (where the radial epicyclic frequency vanishes) in the RP model. 

The accretion discs can be geometrically thin with Keplerian (geodetical) profile of angular velocity \citep{Nov-Tho:1973:BlaHol:,Pag-Tho:1974:ApJ:}, or toroidal, geometrically thick with angular velocity profile governed by gravity and pressure gradients \citep{Koz-etal:1977:ASTRA:,Abr-etal:1978:ASTRA:,Stu-etal:2009:CLAQG:}. Frequency of the disc oscillations is related to the Keplerian (orbital) and epicyclic frequencies of the circular geodesic motion for both Keplerian discs \citep{Kat-Fuk-Min:1998:BHAccDis:,Kat:2004:PASJ:,Now-Leh:1998:TBHAD:} and the slender toroidal discs \citep{Rez-etal:2003:MNRAS:,Mon-Zan:2012:MNRAS:}. The resonance can be of two kinds. The internal, parametric resonance occurs directly between the radial and vertical epicyclic oscillatory modes, giving the basic resonance epicyclic model. The parametric resonance is governed by the Mathieu equation implying strongest resonant phenomena for the frequency ratio $3:2$ \citep{Lan-Lif:1969:Mech:,Nay-Moo:1979:NonOscilations:,Stu-Kot-Tor:2013:ASTRA:} that naturally explain the observed frequency ratio in the twin HF QPOs. On the other hand, the forced non-linear resonance admits presence of combinational (beat) frequencies in the resonant solutions \citep{Nay-Moo:1979:NonOscilations:} -- then the beat frequency $\nu_{-}=\nu_{\theta}-\nu_r$ implies the observed frequency ratio $\nu_{\theta}:\nu_{-}=3:2$ at the radius where the frequency ratio $\nu_{\theta}:\nu_{r}=3:1$, as shown in \cite{Stu-Kot-Tor:2013:ASTRA:}. \footnote{Note that the resonance phenomena could come into the play even in the framework of the hot spot, relativistic precession model \citep{Stu-Kot-Tor:2011:ASTRA:}.} Therefore, we test also the RE1 model where the frequency identification is defined by $\nu_{\rm U}=\nu_{\rm \theta}$ and $\nu_{\rm L}=\nu_{-}$ and the RE2 model with $\nu_{\rm U}=\nu_{-}$ and $\nu_{\rm L}=\nu_\mir$. Defining the beat frequency $\nu_{+}=\nu_{\theta}+\nu_r$, we select the model RE3 with $\nu_{\rm U}=\nu_{+}$ and $\nu_{\rm L}=\nu_\theta$, the model RE4 having $\nu_{\rm U}=\nu_{+}$ and $\nu_{\rm L}=\nu_{-}$, and the model RE5 defined by $\nu_{\rm U}=\nu_\mir$ and $\nu_{\rm L}=\nu_{-}$. Although none of the beat frequency resonance models RE1-5 can explain the twin HF QPOs in any of the three microquasars, it is useful to apply them for the intermediate black hole in the M82 X-1 source in order to test the role of the forced resonant phenomena. 
 
For completeness we add also the WD oscillation model related to the inertial-acoustic modes and the so called g-modes of warped thin disc oscillations \citep{Kat:2004:PASJ:,Kat:2008:PASJ:}, with the identification of the frequencies given by the relations $\nu_{\rm U}=2\nu_{K}-\nu_\mir$ and $\nu_{\rm L}=2(\nu_{K}-\nu_\mir)$ -- this model is not able to explain the twin HF QPO in any of the three microquasars \citep{Tor-etal:2011:ASTRA:}. 

The combinations of the Keplerian and epicyclic frequencies giving the upper and lower frequencies of the twin HF QPOs are presented for all the selected models in Table 1. 

The frequency resonance conditions of the parametric and direct forced resonances are the same, but the physical conditions as the resonant frequency width, resonance strength, and time evolution differ -- for details see \citep{Nay-Moo:1979:NonOscilations:,Lan-Lif:1969:Mech:}. Here we concentrate attention to the resonance frequency conditions only, as the present state of the HF QPO measurements is not precise sufficiently to test the sophisticated details of the predictions of parametric or forced resonances, e.g., the resonance frequency width. We expect that the future generation of the X-ray satellite detectors, as planned in the LOFT observatory \citep{Fer-etal:2012:ExpAstr:}, could make precision of the frequency measurement (in both the central frequency and the QPO peak width) high enough to follow details of the assumed resonance phenomena. 

It should be stressed that details of the resonant phenomena could cause a slight shift from the requirement of strictly rational frequency ratios at resonance. For example, the parametric resonance allows for a scatter of the resonant frequencies, i.e., this kind of resonance can occur while the oscillating modes in resonance have frequency ratio slightly different from the exact rational ratio -- the width of the resonance scatter strongly decreases with increasing order of the resonance \citep{Lan-Lif:1969:Mech:}. In the case of forced resonances, scatter of the frequency ratio is governed by the non-linear effects \citep{Nay-Moo:1979:NonOscilations:}. Therefore, in the case of the M82 X-1 source it is necessary to consider equally all the frequency ratios of the observed twin HF QPOs with $\nu_{\rm U}:\nu_{\rm L}\sim3:2$. 

\begin{figure*}
\centering 
\includegraphics[width=\hsize]{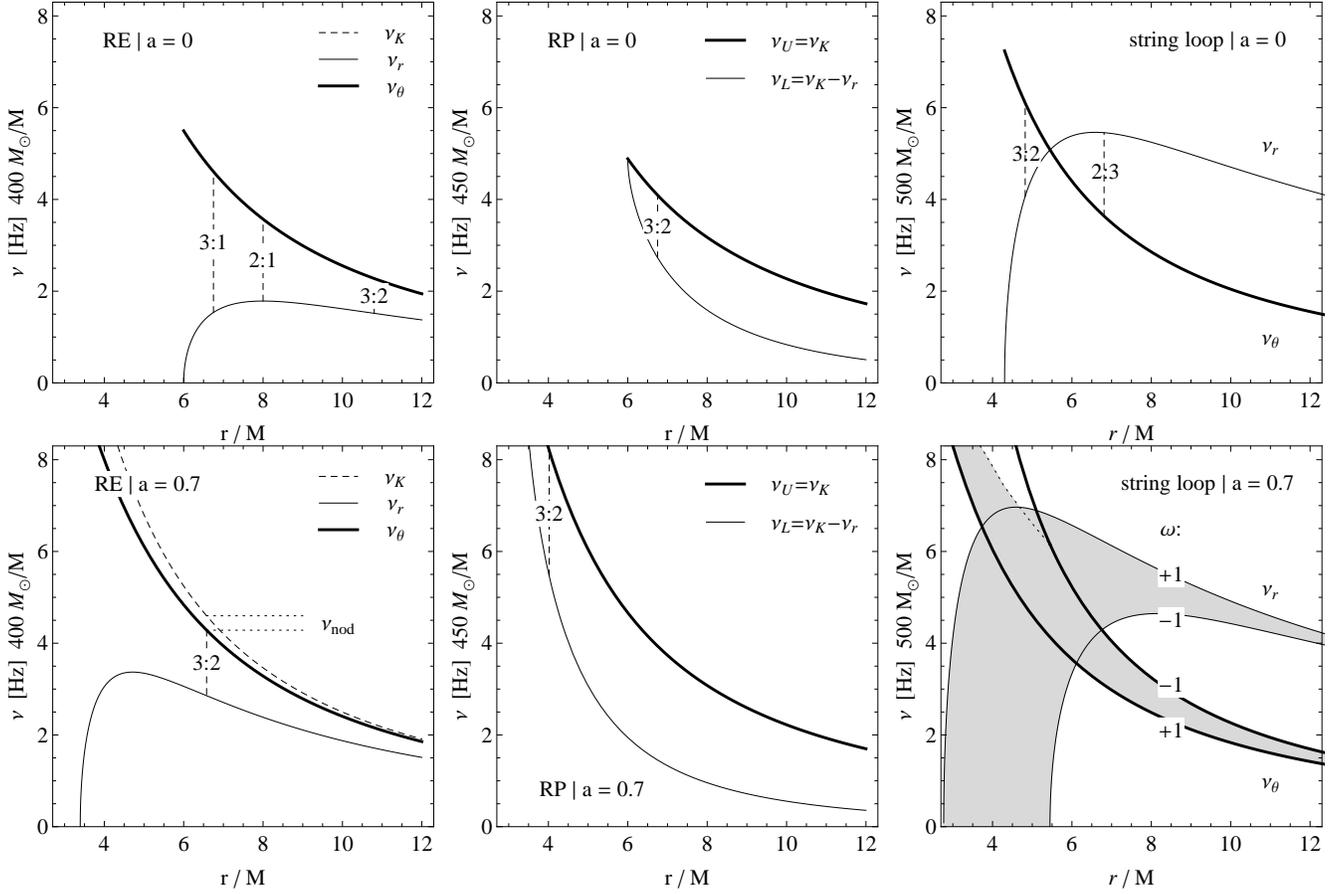}
\caption{
Radial profiles of the Keplerian, radial and latitudinal harmonic frequencies $\nu_\mir(r)$, $\nu_\mir(r)$ and $\nu_\mit(r)$ related to the static distant observers for particle oscillations (first row), their combinations (second row) and string loop frequencies (third row).
\label{RFprofiles}
}
\end{figure*}

\subsection{Frequency of the Keplerian epicyclic motion}\label{geneqcond}

In the Kerr spacetimes characterized by the mass $M$ and dimensionless spin $a$, the circular geodesic motion is restricted to the equatorial plane \citep{Bar-Pre-Teu:1972:ApJ:}. For near-circular epicyclic motion, the vertical epicyclic frequency $\nu_{\theta}$ and the radial epicyclic frequency $\nu_{\rm r}$ take the form \citep{Ali-Gal:1981:GRG:,Kat-Fuk-Min:1998:BHAccDis:,Ste-Vie:1998:ApJ:,Tor-Stu:2005:ASTRA:,Stu-Sche:2012:CLAQG:}
\beq
\label{frequencies}
\nu_{\theta}^2 = \alpha_\theta\,\nu_\mathrm{K}^2, 
\quad
\nu_{\rm r}^2 = \alpha_\mathrm{r}\,\nu_\mathrm{K}^2,
\eeq
where the Keplerian orbital frequency $\nu_\mathrm{K}$ and the corresponding dimensionless epicyclic frequencies are given by the formulae 
\bea
\nu_{\mathrm{K}}&=&\frac{1}{2\pi}\left(\frac{GM}{r_\mathrm{G}^{~3}}\right)^{1/2}
\left(x^{3/2} + a \right)^{-1} =
\frac{1}{2\pi}\left(\frac{c^3}{GM}\right)
\left(x^{3/2}+a\right)^{-1},\nonumber\\
\alpha_\theta&=& 1-4\,a\,x^{-3/2}+3a^2\,x^{-2},
\nonumber\\
\alpha_\mathrm{r}&=&1-6\,x^{-1}+ 8 \,a \, x^{-3/2} -3 \, a^2 \, x^{-2}.
\eea
The dimensionless radius $x = r/(GM/c^2)$, expressed in terms of the gravitational radius $r_G$ of the black hole, is introduced. 

For microquasars, i.e., the black hole low mass X-ray binaries with the observed 3:2 frequency ratio of the observed frequencies of twin HF QPOs, usually the resonance epicyclic model of HF QPOs with the vertical and radial epicyclic oscillations is treated \citep{Tor-etal:2005:ASTRA:}. For a particular resonance n:m of the vertical and radial oscillations, the equation
\beq
\label{ratios} 
\mathrm{n} \, \nu_{\rm r} = \mathrm{m} \, \nu_{\mathrm{\theta}} \, 
\eeq
determines the dimensionless resonance radius $x_{\mathrm{n}:\mathrm{m}}$ as a function of the dimensionless spin $a$ for the direct resonance of the radial and vertical oscillations. This can be easily extended to the resonances with frequencies that combine the Keplerian and epicyclic frequencies -- for details see \citep{Stu-Kot-Tor:2013:ASTRA:}.

We determine radial positions where a given frequency ratio $\nu_U:\nu_L = n:m$ occurs for all considered versions of the geodesic models of HF QPOs with the corresponding combinations of the radial and vertical epicyclic or the~Keplerian oscillation frequencies. In the Kerr black hole spacetimes, the frequencies satisfy the inequality $\nu_{\rm K}>\nu_\mit>\nu_\mir$ for whole the interval $1 \geq a \geq 0$. We use the technique developed in the framework of the resonance models \citep{Stu-Kot-Tor:2013:ASTRA:}, taking into account only the~direct or simple combinational resonances. For all possible resonances, the~resonance condition is given in terms of the~rational frequency ratio parameter
\begin{equation}
p=\left(\frac{m}{n}\right)^2 .
\end{equation}
All the~resonant conditions determining implicitly the~resonant radius $x^{\nu_{\rm U}(K,r,\theta)/\nu_{\rm L}(K,r,\theta)}(a,p)$ have to be related to the~radius of the~innermost stable circular geodesic $x_{\mathrm{ms}}(a)$ giving the~inner edge of Keplerian discs. Therefore, we always require $x^{\nu_{\rm U}(K,r,\theta)/\nu_{\rm L}(K,r,\theta)}(a,p) \geq x_{\mathrm{ms}}(a)$, where $x_{\mathrm{ms}}(a)$ is implicitly given by the relation \citep{Bar-Pre-Teu:1972:ApJ:,Stu-Kot-Tor:2013:ASTRA:}
\begin{equation}
a=a_{\mathrm{ms}}\equiv\frac{\sqrt{x}}{3}\left(4-\sqrt{3x-2}\right).
\end{equation}

It is important to stress that if we assume any two oscillations with given frequencies, having an arbitrary ratio, and occurring at a common radius, the radius can be determined in the same way as if the assumption of the resonant phenomena with exact rational frequency ratio is used. Here we use the generalized condition allowing for the resonance scatter \citep{Nay-Moo:1979:NonOscilations:}, assuming a general, non-rational, ratio of the observed lower and upper frequencies of the twin HF QPOs that is in the vicinity of the $3:2$ ratio as given by the frequency measurement errors presented in \citep{Pas-Str-Mus:2015:Nat:}. Then the~frequency ratio parameter  
\begin{equation}
p = \left(\frac{\nu_{\rm L}}{\nu_{\rm U}}\right)^2 
\end{equation}
determines modified resonance relations for frequency ratios close to the exactly rational ratios. 

In the case of the two basic models, namely the RE and RP models, we give the implicit resonance relations $a=a^{\nu_{\rm U}(K,r,\theta)/\nu_{\rm L}(K,r,\theta)}(x,p)$ for determination of the radius $x^{\nu_{\rm U}(K,r,\theta)/\nu_{\rm L}(K,r,\theta)}(a,p)$ where the twin oscillations given by the frequency ratio $p = (\nu_{\rm L}:\nu_{\rm U})^2$ occur in dependence on the dimensionless spin $a$; the other resonance relations can be found in \citep{Stu-Kot-Tor:2013:ASTRA:}. For the RE model the resonance relation reads 
\beq
a = a^{\mathrm{\theta/r}}(x,p) \equiv \frac{\sqrt{x}}{3(p+1)}\left[ 2(p+2)-\sqrt{(1-p)[3x(p+1)-2(2p+1)]}) \right], \nonumber
\eeq
while for the RP model the relation reads  
\begin{equation}
a=a^{\mathrm{K/(K-r)}}(x,p)\equiv\frac{\sqrt{x}}{3}\left(4-\sqrt{3x(1-p_{\rm RP})-2}\right).
\end{equation}
where
\begin{equation}
p_{\rm RP} = (1 - \sqrt{p})^2.
\end{equation}

In the geodesic models selected for our tests, the upper and lower frequencies of the twin HF~QPOs are related to the combinations of the Keplerian and radial or vertical epicyclic frequencies as presented in the Table I. The nodal frequency is in all the cases determined in the same way as in the standard RP model, i.e., by the relation for the so called Lense-Thirring precession \citep{Mis-Tho-Whe:1973:Gra:}
\bea
 \nu_{\rm nod} = \nu_{\rm K} - \nu_\mit . \label{ff32c}
\eea

\subsection{Fitting the twin HF~QPO frequencies observed in M82~X-1 by the geodesic oscillation models}

Here we restrict our attention to the Kerr black hole case where for any frequency ratio of two oscillations, related to a common radius and governed by the Keplerian and epicyclic frequencies, a unique relation between the radius and the dimensionless spin $a$, $x^{\nu_{\rm U}(K,r,\theta)/\nu_{\rm L}(K,r,\theta)}(a,p)$, exists; frequency of any of the twin HF QPOs then enables to determine uniquely the corresponding relation between the mass of the black hole $M$ and its dimensionless spin $a$, $M^{\nu_{\rm U}(K,r,\theta)/\nu_{\rm L}(K,r,\theta)}(a,p)$, as shown in \citep{Stu-Kot-Tor:2013:ASTRA:}. 

The data of twin HF QPOs observed in the M82 X-1 source display a relatively large error in the frequency measurements \citep{Pas-Str-Mus:2015:Nat:} that do not enable us to determine what kind of resonant phenomena, if any, are at play.
The mean values of $\nu_{\rm U}$ and $\nu_{\rm L}$ are not in exact $3:2$ ratio, but the errors of the frequency measurements do not exclude this ratio and for this reason we are not able to estimate possible shift of the frequency ratio from the exact rational resonant frequency condition corresponding to the resonant frequency width related to the effects of the parametric resonance or the non-linear forced resonant phenomena \citep{Nay-Moo:1979:NonOscilations:,Lan-Lif:1969:Mech:}. \footnote{Precision strong enough for such detailed studies can be expected in data from planned LOFT satellite detectors \citep{Fer-etal:2012:ExpAstr:}.} Therefore, we use in determining the limits on the M82 X-1 black hole mass the simple method of treating the whole observed range of the frequency ratio in the twin HF QPO data for both families of the applied geodesic models of twin HF QPO (hot spot, and resonance). The resonance is not explicitly assumed in both the RP and RE models. Only the frequency ratio is relevant, being taken from whole the interval allowed by the observational data with given errors. The resonance frequency relations are thus applied in the modified form with non-rational frequency ratios implied by the observational errors of the frequency measurements of the twin HF QPOs. Of course, we still assume that both the oscillatory modes are related to a common radius in the Kerr black hole spacetime. 

The method of determining the black hole mass limit implied by the twin HF QPOs is for all the considered geodesic models based on the resonance relations technique developed in \citep{Stu-Kot-Tor:2013:ASTRA:} that is applied for arbitrary, non-rational frequency ratio. We assume that any frequency from the interval of allowed values of the upper frequency can be combined with any frequency from the allowed interval of the lower frequency. The low frequency QPOs are then used to put additional restriction on the black hole parameters. For all the geodesic models of the twin HF QPOs the observed low frequency QPOs are related to the nodal precession with the Lense-Thirring frequency. The resonance relation method thus consists from the following succeeding steps. 

(i) We determine the range of frequency ratios related to the measured upper and lower frequencies of the twin HF QPOs. This range is applied to the ratios of the upper and lower frequencies treated in the framework of a given model. In order to find the limits on the black hole mass, it is enough to consider only the frequency ratios at the edges of the allowed frequency ratio interval because of the character of the radial profiles of the Keplerian and epicyclic frequencies in the Kerr black hole spacetimes \citep{Stu-Kot-Tor:2013:ASTRA:}. 

(ii) We use the frequency resonance relation for the considered geodesic model, $a^{\nu_{\rm U}(K,r,\theta)/\nu_{\rm L}(K,r,\theta)}(x,p)$, and for the maximal and minimal values of the frequency ratio $p$ of the interval given by the observational data we give the related dependence on the radius where the twin oscillations occur. \footnote{In the geodesic models, the frequency ratio relation $\nu_{\rm U}(x;M,a):\nu_{\rm L}(x;M,a)=1/\sqrt{p}$ allows for significant simplification as the mass dependence is cancelled because of the same mass scaling in the Keplerian and epicyclic frequencies \citep{Stu-Kot-Tor:2013:ASTRA:}.} The twin HF QPO radius, $x=x^{\nu_{\rm U}(K,r,\theta)/\nu_{\rm L}(K,r,\theta)}(a,p)$, can be determined for a given dimensionless spin $a$ of the black hole $(0<a<1)$ due to the equation $a=a^{\nu_{\rm U}(K,r\theta)/\nu_{\rm L}(K,r,\theta)}(x,p)$; the radius has to satisfy the condition $x^{\nu_{\rm U}(K,r,\theta)/\nu_{\rm L}(K,r,\theta)}(a,p)>x_{\rm ms}(a)$. Such a solution for radius is unique for the Kerr black holes \citep{Tor-Stu:2005:ASTRA:,Stu-Sche:2012:CLAQG:}. 

(iii) The mass parameter is adjusted by fitting the theoretical, geodesic model upper (or lower) frequency to the corresponding observed frequency. For the chosen upper frequency and the radius $x^{\nu_{\rm U}(K,r,\theta)/\nu_{\rm L}(K,r,\theta)}(a,p)$ related to the spin $a$ by the preceding procedures, we determine the corresponding black hole mass parameter $M^{\nu_{\rm U}(K,r,\theta)/\nu_{\rm L}(K,r,\theta)}(a,p)$ in dependence on the black hole spin $a$. (The same dependence follows from fitting to the lower observed frequency.) This procedure gives the limit on the black hole mass for the whole interval of the allowed black hole spin, $0<a<1$. 

(iv) The restrictions from the nodal frequency model related to the low frequency QPOs are given in the same way as those related to the twin HF QPOs. At each radius predicted by the HF QPOs model under consideration, we assume occurrence of all observed low frequency QPOs equal to the nodal frequency, giving thus the additional limits on the black hole mass. Taking a given value of the spin from the interval $0<a<1$, and the related radius given by the resonance condition of the considered geodesic model, the mass parameter can be determined for the given spin $a$ for each of the nodal frequencies from the observed interval. As in the case of the twin HF QPOs, the lower and upper values of the observed frequencies of the low frequency QPOs are sufficient to give the restrictions introduced by the nodal frequency model. 

(v) Combining restrictions implied by the twin HF QPOs and the low frequency QPOs under assumption of their occurrence at a common radius, we obtain restrictions on the black hole mass and spin for each of the considered geodesic HF QPO models combined with the nodal precession model. The limits on the radius where the simultaneous HF QPOs and low frequency QPOs occur can be obtained using the resonance frequency relation technique. 

The presented technique could be named {\it simultaneous frequency relation technique}, as the precise resonance conditions represent a simple form of this technique, related to the rational frequency ratios. On the other hand, the resonances themselves allow for non-rational ratios, if non-linear phenomena and resonance frequency width enter the play. 

\subsection{Black hole mass and spin limited by the geodesic models of QPOs}

\begin{figure*}
\includegraphics[width=0.325\hsize]{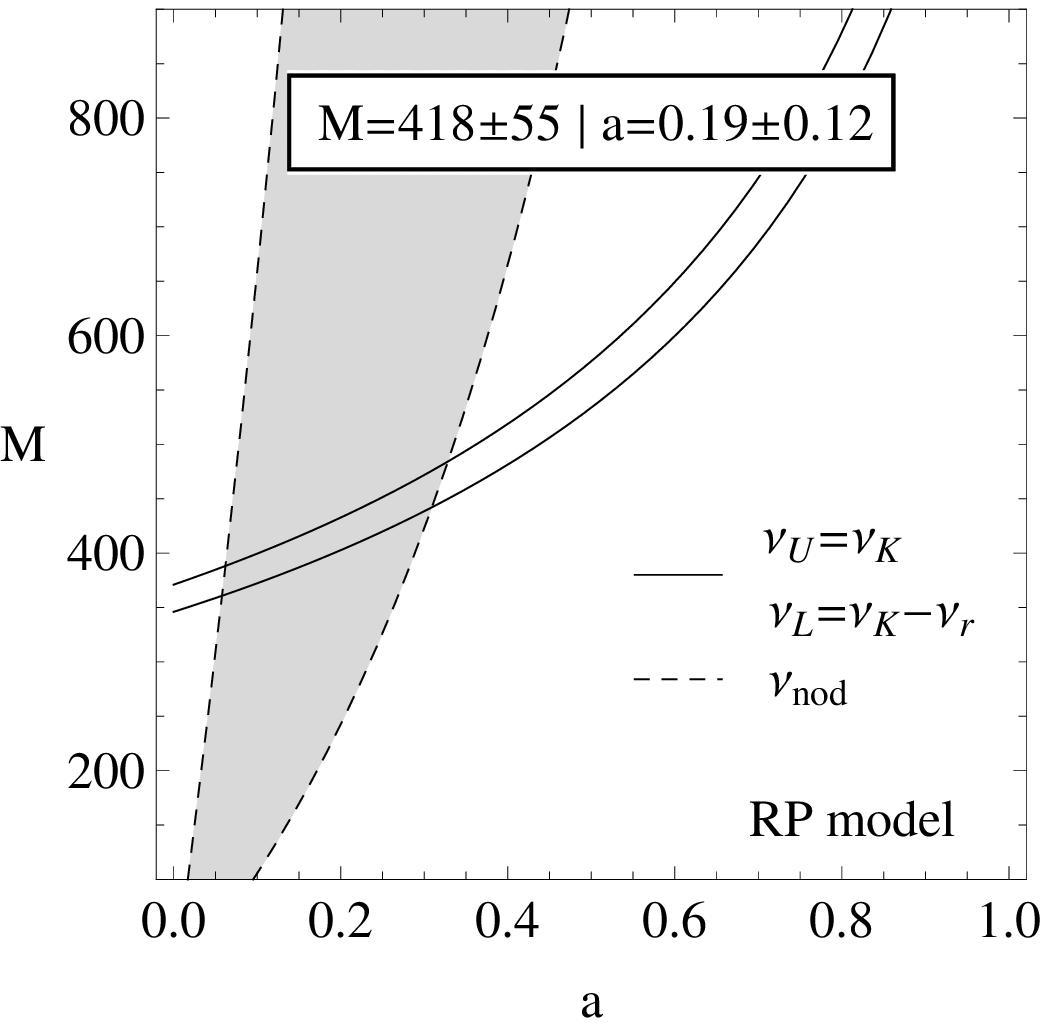}
\includegraphics[width=0.325\hsize]{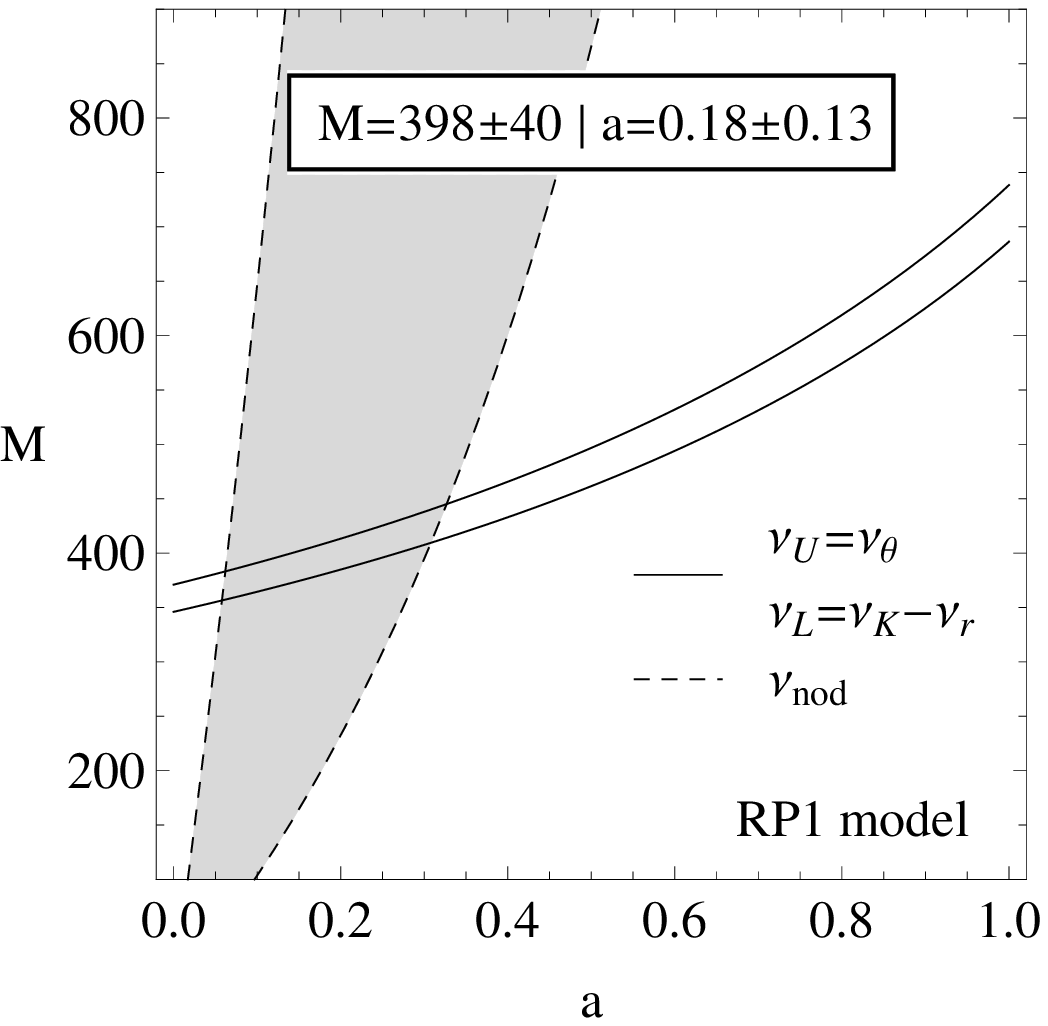}
\includegraphics[width=0.325\hsize]{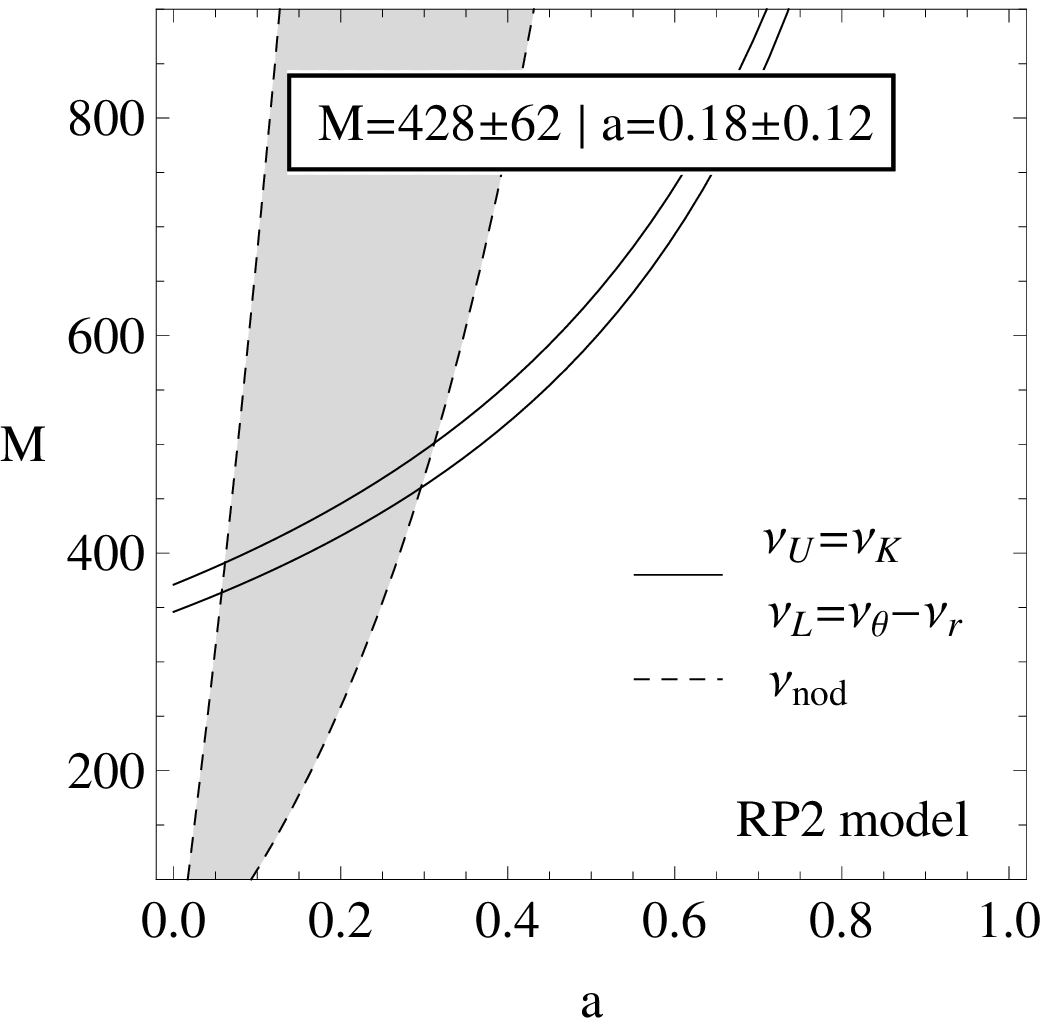}
\caption{
Restrictions on the Kerr spacetime parameters $M$ and $a$ given by the {\it relativistic precession} model and its variants applied to the M82~X-1 source data. 
The solid lines are given by the $3\nu_{\rm L}\sim 2\nu_{\rm U}$ twin HF~QPOs resonance condition giving the $r_{3:2}$ radius with scatter corresponding to the error of measurement of the observed frequencies. The dashed lines are related to the nodal frequency $\nu_{\rm nod}$ for the LF~QPOs assumed at the same $r_{3:2}$ radius - shaded area covers whole the observed LF~QPOs range.
\label{figRP}
}
\end{figure*}

\begin{figure*}
\includegraphics[width=0.325\hsize]{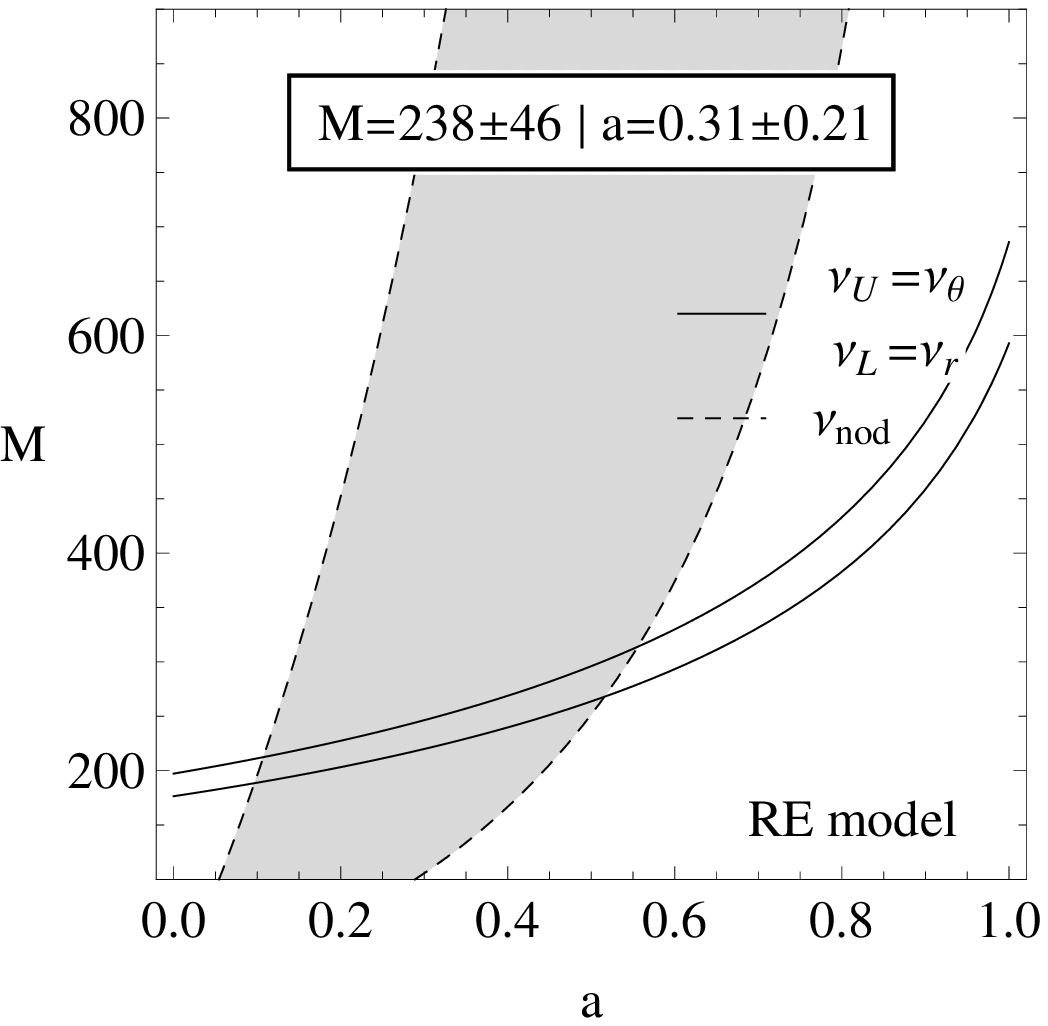}
\includegraphics[width=0.325\hsize]{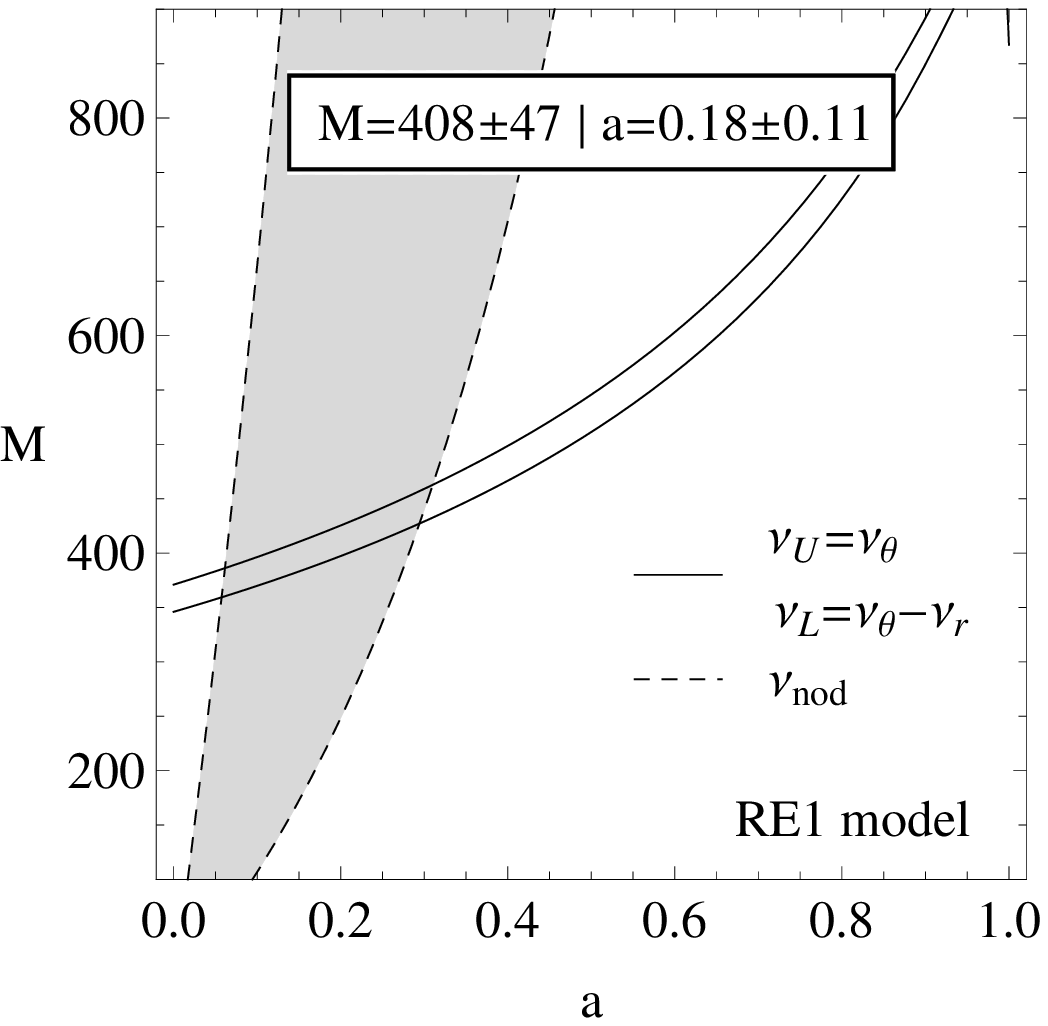}
\includegraphics[width=0.325\hsize]{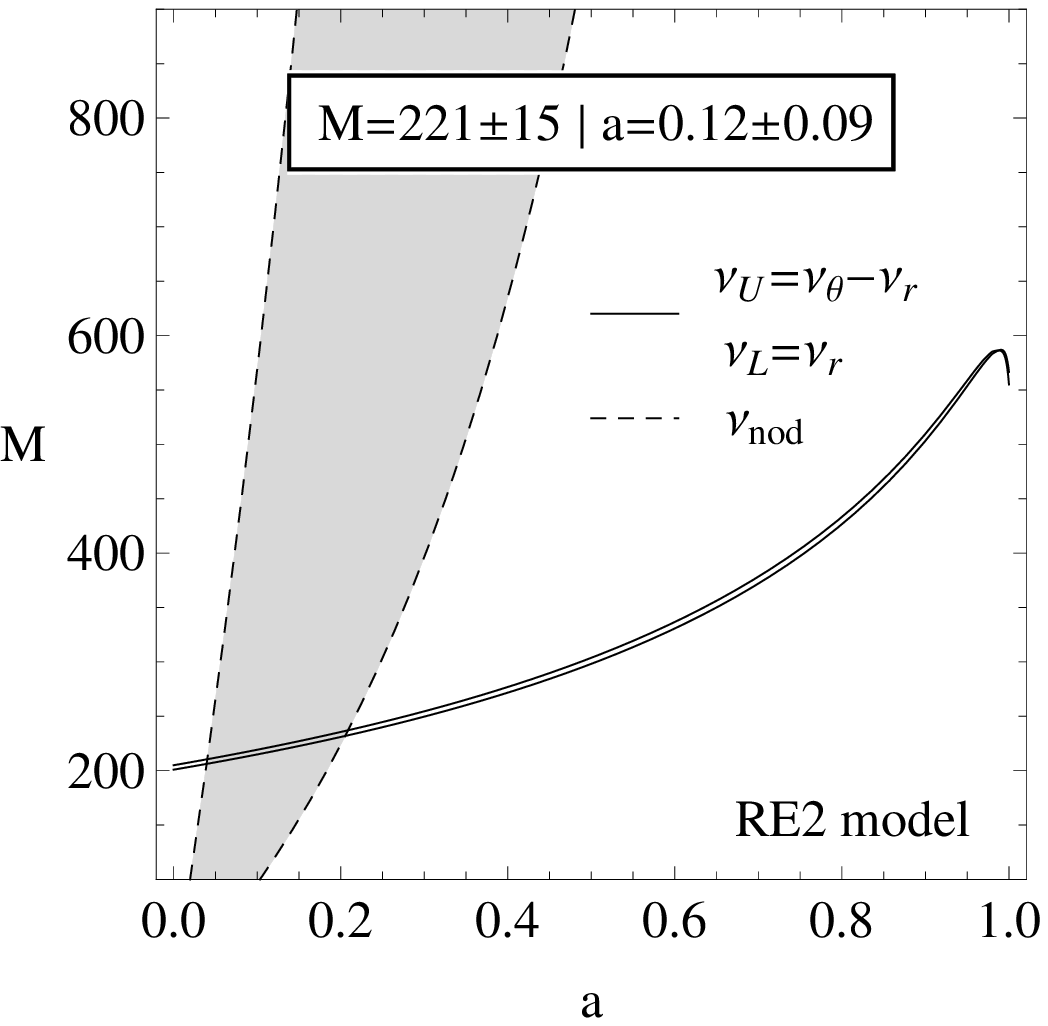}\\
\includegraphics[width=0.325\hsize]{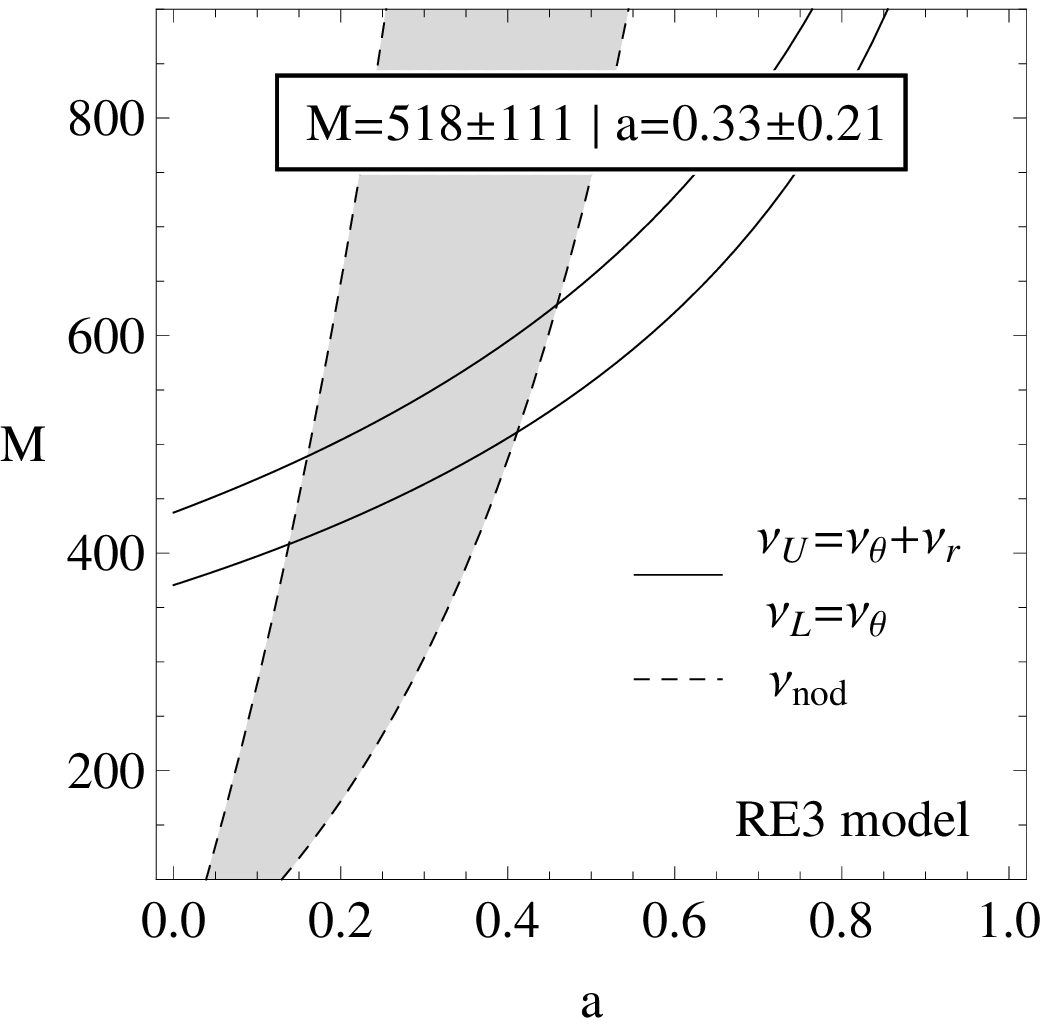}
\includegraphics[width=0.325\hsize]{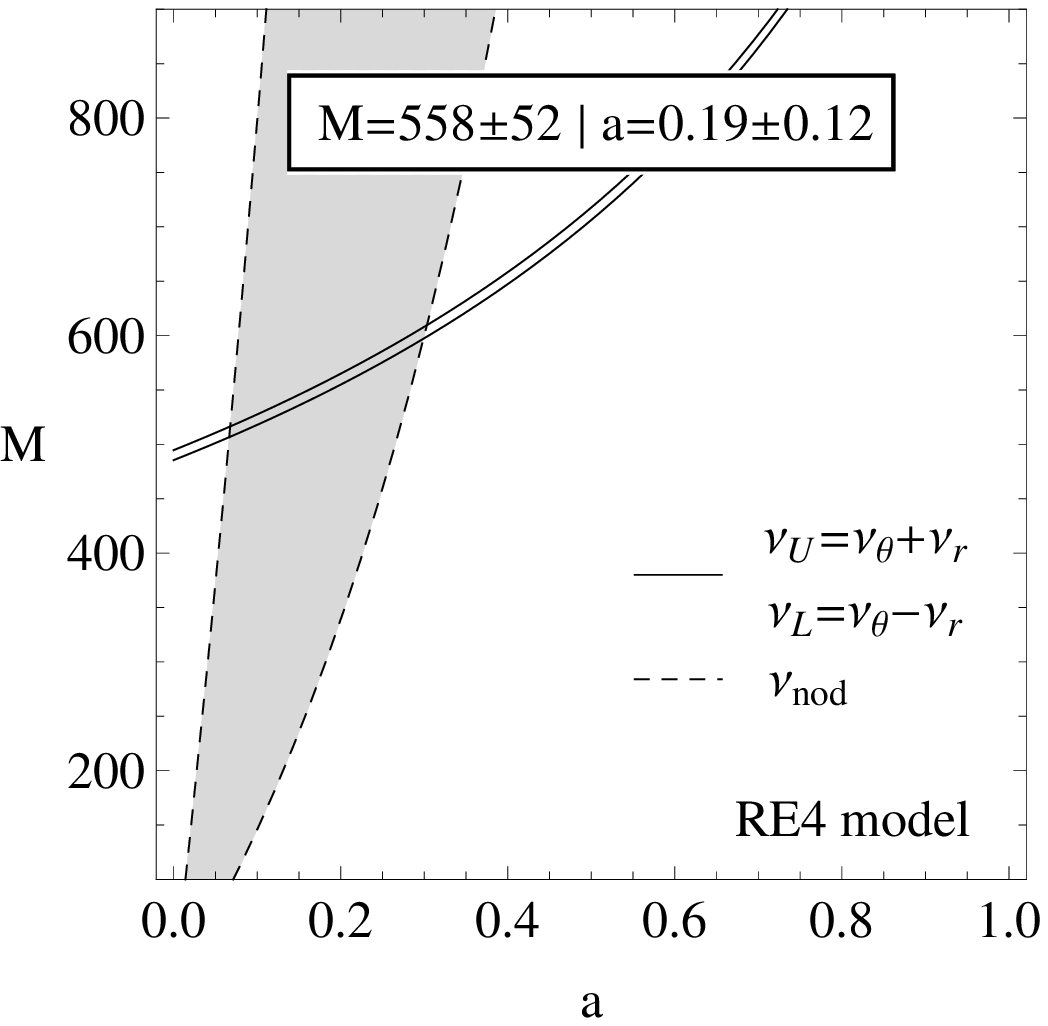}
\includegraphics[width=0.325\hsize]{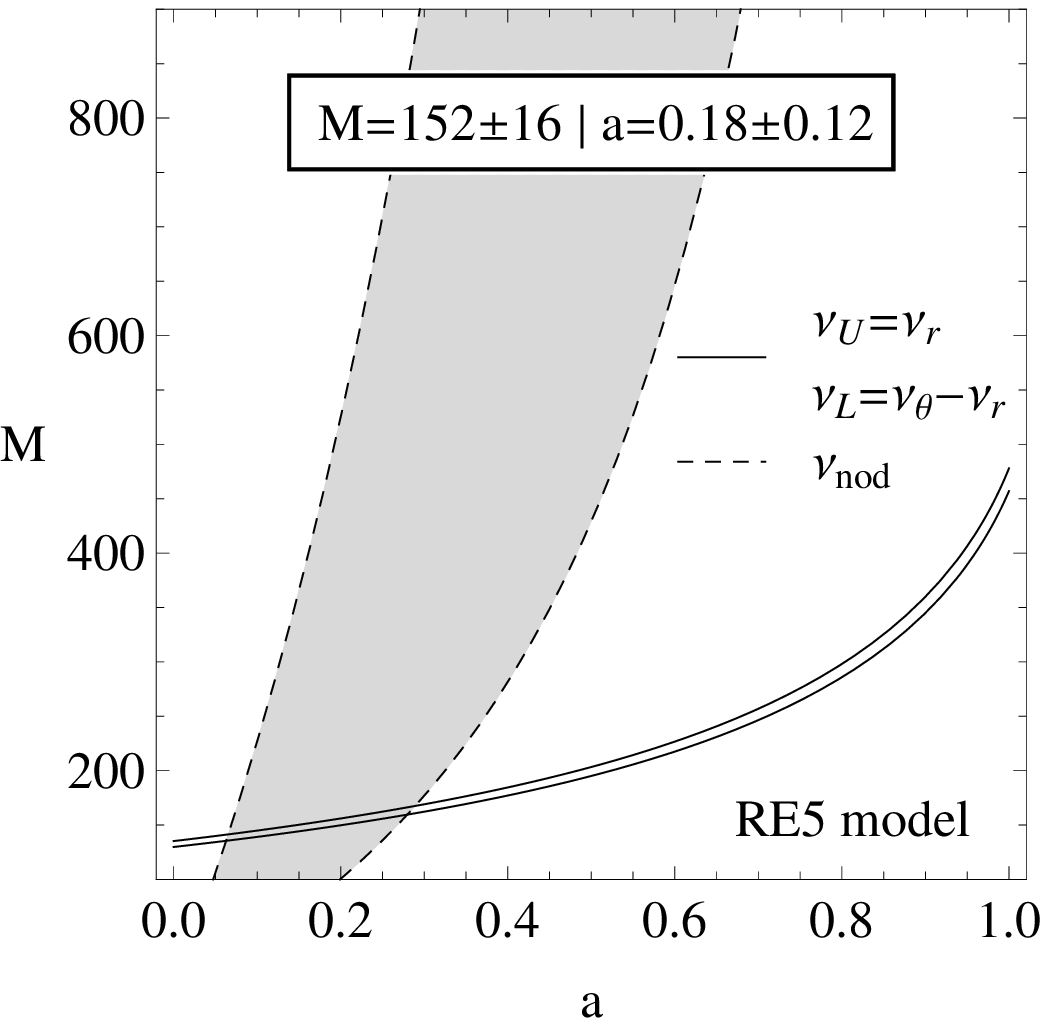}
\caption{
Restrictions on the Kerr spacetime parameters $M$ and $a$ given by the {\it epicyclic resonance} model and its variants applied to the M82~X-1 source data. The method of determining the restrictive areas is the same as in the case of the relativistic precession model. 
\label{figRE}
}
\end{figure*}

\begin{figure*}
\includegraphics[width=0.325\hsize]{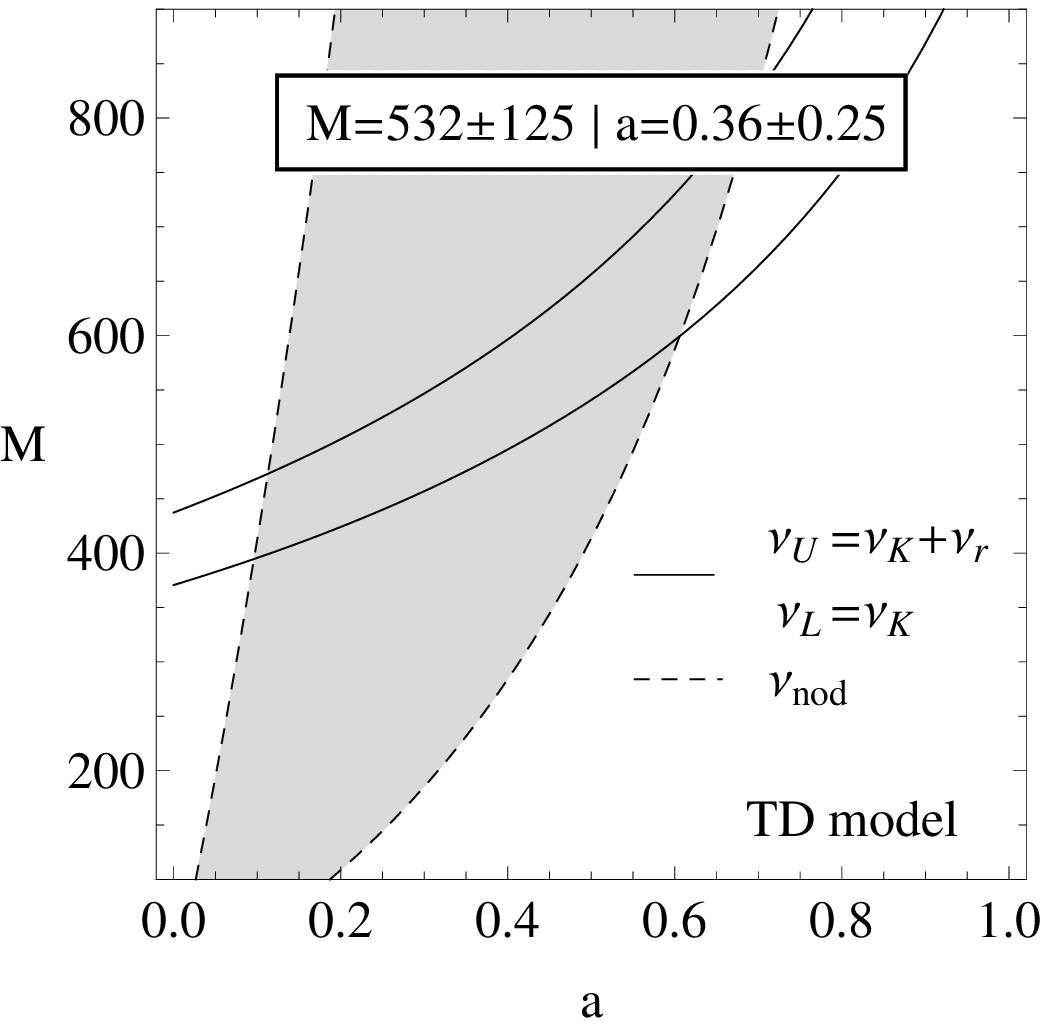}
\includegraphics[width=0.325\hsize]{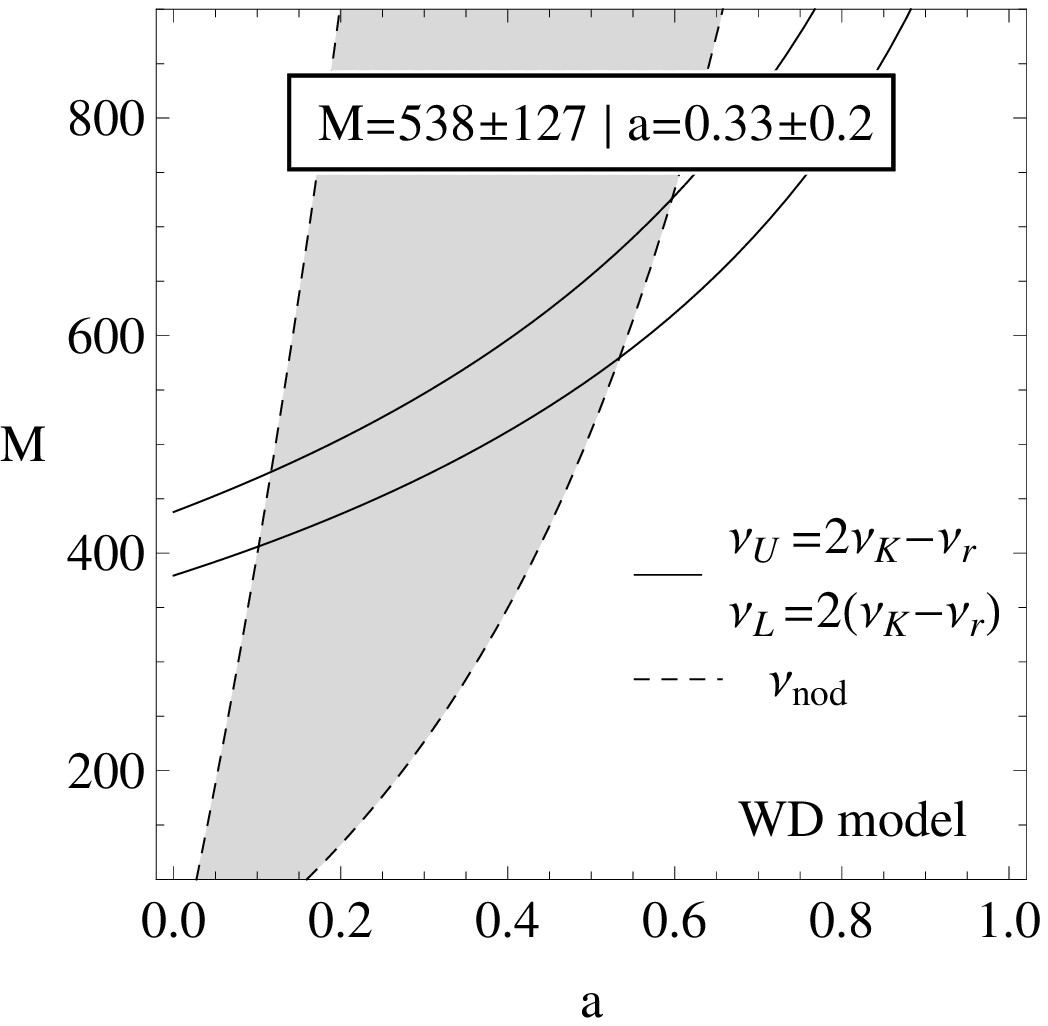}
\caption{
Restrictions on the Kerr spacetime parameters $M$ and $a$ given by the {\it TD} and {\it WD} models applied to  the M82~X-1 source data. The method of determining the restrictive areas is the same as in the case of the relativistic precession model. 
\label{figTDTW}
}
\end{figure*}

Results of the numerical calculations are presented in Figure \ref{figRP} for the RP model and its variants RP1 and RP2, in Figure \ref{figRE} for the RE model and its variants RE1 and RE2, and in Figure \ref{figTDTW} for the special cases of the TD model and the WD model. The frequency relations governing the upper and lower HF QPOs in the selected geodesic models are presented in Table \ref{tab1}, where the range of the allowed values of the black hole mass, $M_{\rm min}$--$M_{\rm max}$, determined by the selected geodesic models of twin HF QPOs for the range of the black hole spin, $0<a<1$, is presented. Finally, the intervals of the black hole mass and spin, allowed by the combined restrictions of the twin HF QPOs geodesic models and the nodal frequency model of the low frequency QPOs are given in Table 1. 

We can see that the restrictions are strongly model dependent. In the case of the RP model and its variants, the lower limit of the twin HF QPOs is common, $M_{\rm min}=350~M_{\odot}$, but the upper limit can be as large as $M_{\rm max}=1500~M_{\odot}$ in the case of the RP model. When the nodal frequency restrictions are added, all the RP models predict the black hole mass $M\sim400~M_{\odot}$ and the spin $a\sim0.2$. 
The RE models predict lower restrictions in comparison to those given by the RP models, $M_{\rm min}\sim150~M_{\odot}$,  $M_{\rm max}\sim700~M_{\odot}$, but the lower restriction of the RE1 model, using the beat frequency, is nearly the same as those of the RP models. 
The combined restrictions due to the RE model predict the black hole mass $M\sim240~M_{\odot}$ and the spin $a\sim0.31$, while the RP2 model predicts $M\sim220~M_{\odot}$, but the spin $a\sim0.12$. The RE1 model predicts combined restrictions similar to those of the RP models. The RE3 model predicts large mass $M\sim520~M_{\odot}$ and large spin $a\sim0.33$, while the RE4 model predicts the largest mass $M\sim560~M_{\odot}$ and lower spin $a\sim0.19$, but the RE5 variant predicts the lowest mass $M\sim150~M_{\odot}$ and spin $a\sim0.18$.
The TD and WD models predict almost identical limits of the mass and spin that are slightly higher than those of the RP models, $M\sim540~M_{\odot}$, $a\sim0.35$. In all the considered variants of the geodesic model, the common radius of the twin HF QPOs and the related low frequency QPO has to be lower than $\sim 10M$, i.e., it is located closely to the inner edge of the Keplerian disc. 
Except the case of the predictions of the RE5 model, all the other geodesic oscillation models give mass restrictions that are in agreement with the limit of $200 - 800 M_{\odot}$ implied by the spectral measurements \citep{Fen-Kaa:2010:ApJ:}. 

Note that the RP model mass range implied by the resonance relation technique, $M\sim418\pm55~M_{\odot}$, agrees well with the mass range $M\sim415\pm63~M_{\odot}$ implied by the Monte Carlo technique \citep{Pas-Str-Mus:2015:Nat:,Mot-etal:2014:MNRAS:}. 

\section{String loop oscillation model}

The axisymmetric current-carrying string loops, governed by tension preventing their expansion behind some radius and by angular momentum that prevents them from collapse, can oscillate around stable equilibrium radii at the equatorial plane of the Kerr black holes, giving rise to observational phenomena related to the HF QPOs \citep{Jac-Sot:2009:PHYSR4:,Kol-Stu:2010:PHYSR4:,Stu-Kol:2012:PHYSR4:,Kol-Stu:2013:PHYSR4:}. It is important that the so called transmutation effect related to the string loops, i.e., transmission of their oscillatory internal energy into energy of the translational motion, can be also astrophysically interesting, as it causes an outward-directed acceleration of the string loops in the gravitational field of compact objects -- neutron stars or black holes \citep{Lar:1994:CLAQG:,Jac-Sot:2009:PHYSR4:,Stu-Kol:2012:PHYSR4:,Stu-Kol:2012:JCAP:,Kol-Stu:2013:PHYSR4:}. Such an effect can be important also for electrically charged axisymmetric string loops moving in combined external gravitational and electromagnetic fields \citep{Lar:1993:CLAQG:,Tur-etal:2013:PHYSR4:,Tur-etal:2014:PHYSR4:}. Acceleration of the string loops in the deep gravitational field of black holes can be extremely efficient, leading to ultra-relativistic escaping velocities of the string loops \citep{Stu-Kol:2012:JCAP:,Kol-Stu:2013:PHYSR4:}, only slightly dependent on the black hole spin \citep{Kol-Stu:2013:PHYSR4:}. Therefore, the string loop transmutation effect can serve as an alternative model of the formation and collimation of the ultra-relativistic jets in active galactic nuclei or Galactic microquasars. For jets escaping from active galactic nuclei the cosmic repulsion can be also relevant behind the so called static radius \citep{Stu-Kol:2012:PHYSR4:,Stu:1983:BULAI:,Stu-Hle:1999:PHYSR4:}. The high-energy, escaping string loops can be thus considered as a model of relativistic jets, while the low-energy, oscillating string loops trapped in the vicinity of the black hole horizon can serve as a model of twin HF QPOs. 

\subsection{Stationary radii of axisymmetric string loops}

Dynamics of the string loops can be effectively described by a properly defined Hamiltonian governing formally, in correspondence to the test particle case, motion of one point of the axisymmetric string loop in axisymmetric external fields \citep{Kol-Stu:2013:PHYSR4:,Tur-etal:2014:PHYSR4:}. The dynamics of the string loops is then governed by the parameters of the external gravitational field, and two parameters, $J, \omega$, governing the combined effects of the string tension and angular momentum. The string loop dynamics can be determined by an effective potential $E_{\rm b}(r;a,J,\omega)$ related to their energy parameter $E$. It is called energy boundary function and gives the turning points of the radial motion of the string loop \citep{Stu-Kol:2014:PHYSR4:}.

The stationary points, i.e., the local extrema of the energy boundary function $E_{\rm b}(r;a,J,\omega)$, governing the equilibrium positions of the string loops in the equatorial plane of Kerr black holes ($\theta=\pi/2$), are determined by the function $J^2_{\rm E}(r;a,\omega)$ defined by the relation \citep{Kol-Stu:2013:PHYSR4:,Stu-Kol:2014:PHYSR4:}
\beq
J^2_{\rm E}(r;a,\omega) =  \frac{(r-1) \left(\omega^2+1\right) H^2}{4 a \omega \sqrt{\Delta} \left(a^2+3 r^2\right)+\left(\omega^2+1\right) F }, \label{eqrovina}
\eeq
where 
\beq
 \fl H(r;a) = a^2 (r+2) +r^3, \quad F(r;a) = (r-3) r^4 -2 a^4+a^2 r \left(r^2-3 r+6\right). \label{Ffce}
\eeq

A detailed discussion of the properties of the energy boundary function $E_{\rm b}(r;a,J,\omega)$ and the string loop motion can be found in \citep{Kol-Stu:2013:PHYSR4:,Stu-Kol:2014:PHYSR4:}. We have to concentrate on the situation when for a string loop with fixed values of the angular momentum parameters $J$ and $\omega$, closed $E={\rm{}const}$ sections of the effective potential (energy boundary function) occur around a stable equilibrium position of the string loop being given by the equation 
\bea
              J^2 = J^2_{\rm E}(r;a,\omega) .
\eea
The stable equilibrium positions correspond to the minimal energy related to the string loop with the angular momentum parameters $J,\omega$. Around such stable equilibrium positions, small oscillations of string loops occur, if their energy slightly exceeds the minimal value. The analysis of the oscillatory motion of string loops around their stable equilibrium positions, using the perturbative treatment of the Hamiltonian, can be found in \citep{Stu-Kol:2012:JCAP:,Kol-Stu:2013:PHYSR4:,Stu-Kol:2014:PHYSR4:}. 

Small oscillations of the string loops around stable equilibrium positions in the equatorial plane of the Kerr geometry represent, in the lowest approximation of the Taylor expansion of the Hamiltonian around any stable equilibrium position of the string loop, two uncoupled linear harmonic oscillators governing the radial and vertical oscillations of the string loop; the higher-order terms govern non-linear phenomena and subsequent transition to the quasi-periodic and chaotic oscillatory motion \citep{Kol-Stu:2013:PHYSR4:}. The frequencies of the radial and vertical harmonic oscillations of the string loops are relevant also in the quasi-periodic stages of the motion and their radial profiles were given and discussed in \citep{Stu-Kol:2014:PHYSR4:}. The frequencies of the string loop harmonic or quasi-harmonic oscillations can fit frequencies of the HF QPOs observed with frequency ratio $3:2$ in three Galactic microquasars GRS~1915+105, XTE~1550-564, GRO~1655-40, i.e, Low-Mass X-ray Binary (LMXB) systems containing a black hole \citep{Stu-Kol:2014:PHYSR4:}. 

\begin{figure*}
\includegraphics[height=0.325\hsize]{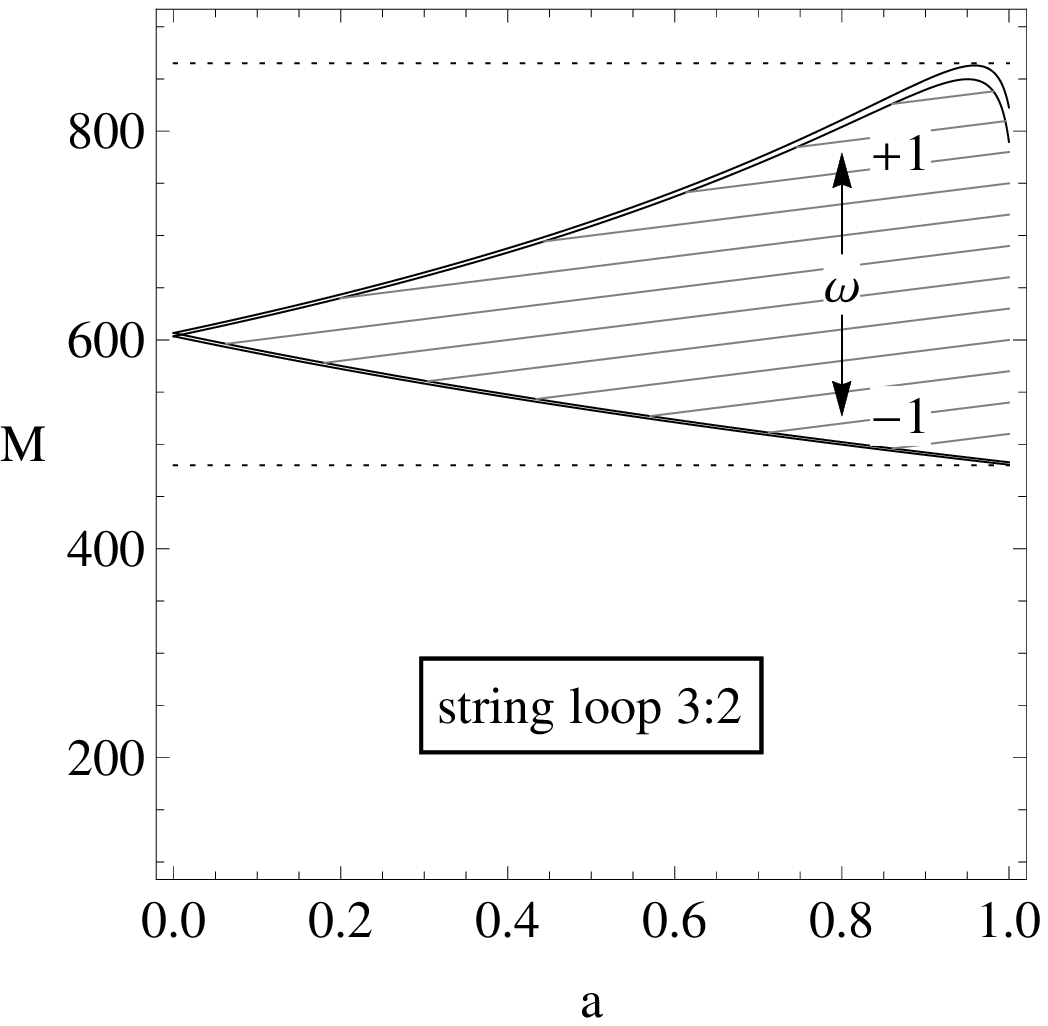}
\includegraphics[height=0.325\hsize]{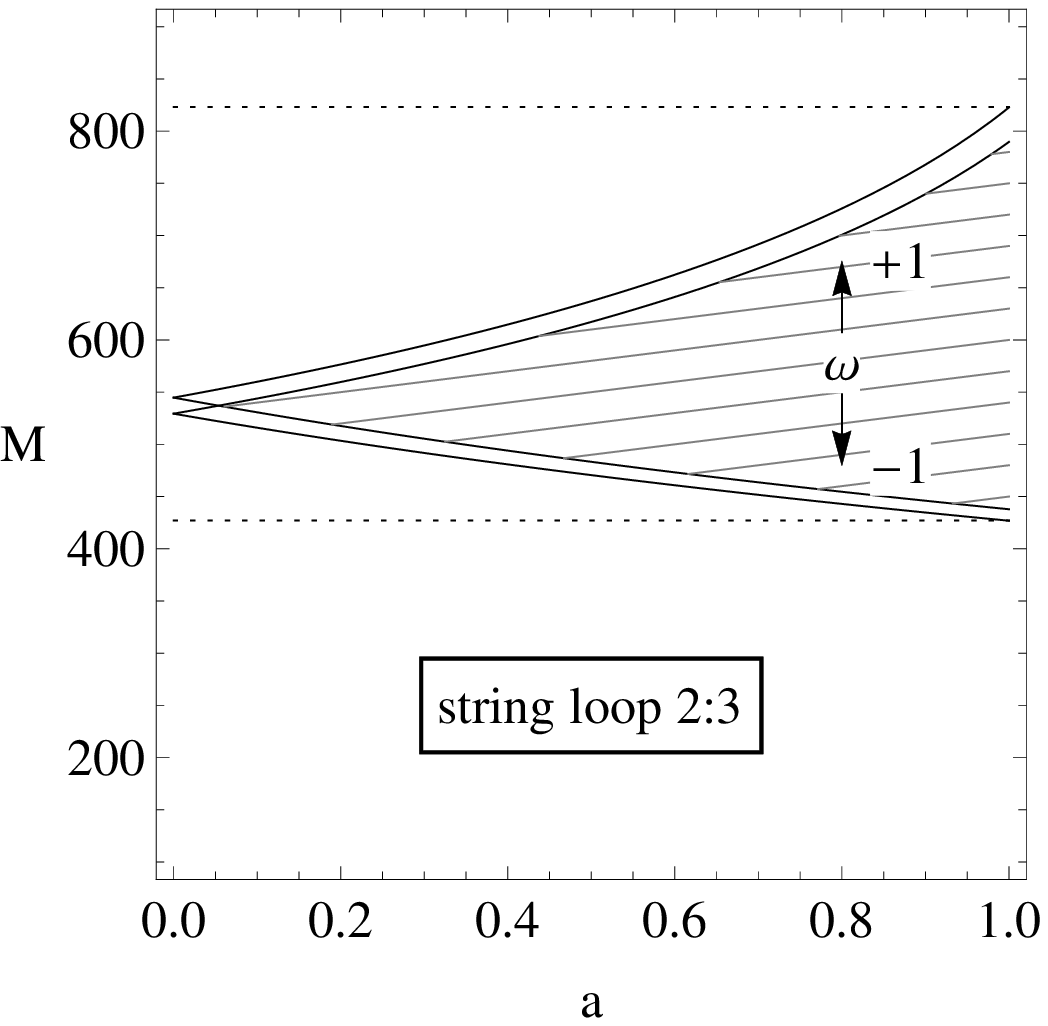}
\caption{
Restrictions on the Kerr spacetime parameters $M$ and $a$ given by the {\it string loop} model applied to the M82~X-1 source data. 
The solid lines are given by $3\nu_{\rm r} \sim 2\nu_{\rm \theta}$ and $2\nu_{\rm r} \sim 3\nu_{\rm \theta}$ twin HF~QPOs resonance condition - hatched  area covers whole the $\omega\in\langle-1,1\rangle$ range. In this case whole the range of the black hole spin, $0<a<1$, is considered and the restriction is put on the mass parameter only. 
\label{figLOOP}
}
\end{figure*}

\subsection{Frequency of the string-loop radial and vertical oscillatory modes}

For the radial and latitudinal (vertical) harmonic oscillatory string loop motion in the Kerr spacetimes, the frequencies related to distant observers are given by \citep{Stu-Kol:2014:PHYSR4:} 
\beq
      \nu_{r} = \frac{c^3}{2\pi GM} \, \Omega_{r}, \quad \nu_{\theta} = \frac{c^3}{2\pi GM} \, \Omega_{\theta},
\eeq
where the dimensionless angular frequencies read
\bea
\fl \Omega^2_{\mir}(r) &=& \frac{ J_{\rm E(ex)} \, \left(2 a \omega \sqrt{\Delta } \left(a^2+3 r^2\right)+\left(\omega ^2+1\right) \left(a^2 r^3 -a^2 \Delta  +r^5-2 r^4\right)\right)}{2 r \left(a^2
   (r+2)+r^3\right)^2 \left(2a\omega  \left(a^2+3r^2\right)+\sqrt{\Delta } \left(\omega ^2+1\right) \left(r^3-a^2\right)\right)^2} , \\
\fl \Omega^2_{\mit}(r) &=& 
\frac{2 a \omega \sqrt{\Delta }  \left(2 a^2 -3 a^2 r -3 r^3\right)+\left(\omega ^2+1\right) \left(a^4 (3 r-2)+2 a^2 (2 r-3) r^2+r^5\right)}
{ r^2 \left(a^2 (r+2)+r^3\right) \left(2a\omega \left(a^2+3r^2\right) \Delta^{-1/2} + \left(\omega ^2+1\right) \left(r^3-a^2\right)\right)},
\eea
with
\bea
\fl J_{\rm E(ex)} (r) &\equiv& H \left(\omega ^2+1\right) (r-1) \left( 6 a^2 r -3 a^2 r^2 -6 a^2 -5 r^4+ 12 r^3 \right) \nonumber \\
\fl && + \left(\omega ^2+1\right) \left[2 F (a^2+3r^2) (1-r) - F H \right] + 8 a \omega \sqrt{\Delta} (r-1) (a^2+3r^2)^2  \nonumber \\
\fl && + 4 a \omega \Delta^{-1/2} H \left[ (a^2+3r^2) \left(\Delta -(r-1)^2\right) -6 r \Delta (r-1) \right]. \label{JEex} 
\eea
The function $J_{\rm E(ex)}(r;a,\omega)$ governs the local extrema of the function $J_{\rm E}(r;a,\omega)$. Its zero points determine the marginally stable equilibrium positions of the string loops - at the zero points the frequency of the radial oscillatory modes of the string loops vanishes. \footnote{Exactly the same situation occurs in the case of the radial epicyclic motion of test particles \citep{Tor-Stu:2005:ASTRA:}.} The string-loop oscillations are possible only if the stable equilibrium positions of the string loops are allowed. The conditions 
\beq
  J_{\rm E(ex)}=0 \quad \textrm{and} \quad J_{\rm E}^2 \geq 0,
\eeq   
satisfied simultaneously, put the limit on validity of the formulae giving the angular frequencies of the radial and vertical oscillations - for details see \citep{Stu-Kol:2014:PHYSR4:}. 

The radial profiles of the string loop oscillations qualitatively differ from those related to the radial and vertical oscillations of the geodesic, test particle motion in the Kerr geometry. There is a crossing point of the radial profiles of the radial and vertical frequencies of the string loop oscillations in the Kerr black hole spacetimes \citep{Stu-Kol:2014:PHYSR4:}. This property of the radial profiles of the radial and vertical frequencies of the string loop oscillations introduces an unambiquity into the fitting of the HF QPOs even in the black hole spacetimes and makes the estimates of the black hole mass more complex. However, it can be conveniently applied to explain the special frequency set of HF QPOs observed recently in the source XTE J1701-407 that is a LMXB system containing a neutron star \citep{Paw-etal:2013:MONRAS:,Stu-Kol:2015:GRG:}. The radial profiles of the string loop oscillations are demonstrated in Figure \ref{RFprofiles} for value of the Kerr spin parameter $a=0.7$ and in the \Schw{} spacetime ($a=0$), where the situation is degenerate, since both the frequencies are independent of the stringy parameter $\omega$. 

\subsection{Fitting the twin HF QPO frequencies observed in M82 X-1 by oscillating string loops}

We assume relevance of the resonance phenomena in the string loop oscillatory motion that are governed by the Kolmogorov-Arnold-Moser theory \citep{Arnold:1978:book:}, as the frequency ratio $\sim 3:2$ is demonstrated at the twin HF QPOs observed in the M82~X-1 source. Therefore, we consider the frequencies with ratios $\nu_{\mit}:\nu_{\mir} \sim 3:2$, or $\nu_{\mit}:\nu_{\mir} \sim 2:3$, to be directly related to the observed values of the QPO frequencies in the source. We identify directly the frequencies $\nu_{\rm U}, \nu_{\rm L}$ with $\nu_{\mit}, \nu_{\mir}$ or $\nu_{\mir}, \nu_{\mit}$ frequencies. 

The procedure of fitting the string loop oscillation frequencies to the observed frequencies has been developed by \cite{Stu-Kol:2014:PHYSR4:} and is presented in Figure \ref{figLOOP} for both cases $3:2$ and $2:3$ of possible combinations of the resonant radii of the string loop oscillations. The fitting procedure then gives for the observed events an allowed region of the parameter space of the spacetime parameters $M,a$, determined by the limiting values of the string loop parameter $\omega\in\langle-1,1\rangle$. Due to the degeneracy of the radial profiles of the string loop oscillation frequencies in the \Schw{} spacetimes ($a=0$), i.e., their independence of the stringy parameter $\omega$, the fitting predicts only one value of the mass parameter $M$ for the spin $a=0$ at each observational event. Extension of the allowed region related to the whole interval of string loop parameter $\omega\in\langle-1,1\rangle $ (i.e., the interval of allowed values of $M$) increases with increasing spin $a$. Therefore, the string loop oscillation model implies a triangular - "carrot" like limit on the spacetime parameters $M,a$ for each of the observed HF QPOs -- see Figure \ref{figLOOP}. The limiting values of the black hole mass are presented in Table 1. They are clearly close to the limiting values predicted by the RP model and its variants. In the case of the string loop oscillation model the additional nodal frequency model is not relevant, and is not applied for the mass estimates. For possibility to obtain a low frequency string loop oscillations see \citep{Stu-Kol:2015:GRG:}

\begin{table*}
\begin{center}
\begin{tabular}{l l l l l l l l }
\hline
model & LMXB & $\nu_U$ 					& $\nu_L$  & $M_{\rm min}$--$M_{\rm max}/M_{\odot}$ & $M/M_{\odot}$ & $a$ & $r_{3:2}$\\
\hline \hline
 RP  & 1655 &$\nu_{\rm K}$ 			& $\nu_{\rm K}-\nu_r $ 		& 346--1505 & $418\pm55$ & $0.19\pm0.12$ & $6.2\pm0.5$ \\
 RP1 & 1550 &$\nu_{\rm \theta}$ & $\nu_{\rm K} - \nu_r $  & 346--738  & $398\pm40$ & $0.18\pm0.13$ & $6.2\pm0.5$ \\
 RP2 & 1655 &$\nu_{\rm K}$ 			& $\nu_\theta -\nu_r $ 		& 346--1400 & $428\pm62$ & $0.18\pm0.12$ & $6.1\pm0.6$ \\
\hline 
 RE  &  1915 &$\nu_\theta$ 							& $\nu_r $ 									& 176--686  & $238\pm46$ & $0.31\pm0.21$ & $8.9\pm1.4$ \\
 RE1 & None &$\nu_{\rm \theta}$ 				& $\nu_{\rm \theta}-\nu_r $ & 346--983  & $408\pm47$ & $0.18\pm0.11$ & $6.2\pm0.5$ \\
 RE2 & None &$\nu_{\rm \theta}-\nu_{r}$ & $\nu_r $ 									& 201--586  & $221\pm15$ & $0.12\pm0.09$ & $6.6\pm0.4$ \\
 RE3 & None &$\nu_{\rm \theta}+\nu_{r}$ & $\nu_\theta $ 						& 371--1448 & $518\pm111$& $0.33\pm0.21$ & $7.0\pm1.3$ \\
 RE4 & None &$\nu_{\rm \theta}+\nu_{r}$ & $\nu_\theta-\nu_{r} $ 		& 486--1176 & $558\pm52$ & $0.19\pm0.12$ & $5.6\pm0.5$ \\
 RE5 & None &$\nu_r $ 								 	& $\nu_{\rm \theta}-\nu_{r}$& 130--478  & $152\pm16$ & $0.18\pm0.12$ & $8.6\pm0.7$ \\
\hline
 TD  & None &$\nu_{\rm K} + \nu_r$ 			& $\nu_{\rm K}$ 						& 371--1407 & $532\pm125$ & $0.36\pm0.25$ & $7.1\pm1.3$ \\
 WD  & None &$2\nu_{\rm K} - \nu_r$ 		& 2($\nu_{\rm K}-\nu_r) $		& 379--1390 & $538\pm127$ & $0.33\pm0.20$ & $6.9\pm1.0$\\
\hline
 string loop 3:2 & All &$\nu_\theta$ & $\nu_r$ 		 & 480--865 & & & (1.6--6.5) \\
 string loop 2:3 & All &$\nu_r$ 			& $\nu_\theta$ & 427--823 & & & (3.8--9.1) \\
\hline
\end{tabular}
\caption{
Restrictions on the parameters $M$ and $a$ of the intermediate mass black hole in the M82~X-1 source are given for various considered QPO models (see section \ref{GOmodels}) for radial $\nu_r$, vertical $\nu_\theta$ and Kepplerian $\nu_{\rm K}$ harmonic frequencies. The 'LMXB' column indicates whether the model has been successful in explaining the 3:2 twin HF QPOs in three microquasars GRS 1915+105, XTE 1550-564 and GRO 1655-40.
\label{tab1}
} 
\end{center}
\end{table*}

\section{Conclusions}

We have applied a variety of geodesic models of the twin HF QPOs, namely the RP and RE models, on the twin HF QPO frequency set reported quite recently for the extragalactic M82 X-1 source \citep{Pas-Str-Mus:2015:Nat:}, in order to put the limits on the black hole assumed in this source and test the expectation that this is an intermediate black hole. In Table \ref{tab1}. we summarize the models that have been successfully applied to the HF QPOs in the stellar-mass black hole binary systems.
We have used the method of resonance relations introduced in \citep{Stu-Kot-Tor:2013:ASTRA:}. The restrictions given by the geodesic models of the twin HF QPOs were combined with those given by the nodal frequency model of the low frequency QPOs observed in the M82~X-1 source. 

For comparison, we have used also the recently introduced (non-geodesic) string loop oscillation model that can be well applied to the frequencies of the twin HF QPOs observed in Galactic microquasars \citep{Stu-Kol:2014:PHYSR4:}, or some neutron star LMXBs \citep{Stu-Kol:2015:GRG:}. 

The limits on the mass and spin of the M82 X-1 black hole implied by the twin HF QPOs combined with those of the low frequency QPOs in the framework of the standard RP model including the nodal precession are similar for the method of the resonance relations \citep{Stu-Kot-Tor:2013:ASTRA:} used in the present paper, and the Monte Carlo method introduced in \citep{Mot-etal:2014:MNRAS:} -- the mass range given by the resonance relation method, $M\sim418\pm55~M_{\odot}$, is just contained in the mass range given by the Monte Carlo method, $M\sim415\pm63~M_{\odot}$, indicating applicability of both the methods for rough estimates of admitted mass in the sources demonstrating QPOs. 

We can conclude that the restrictions on the M82~X-1 black hole mass are strongly model dependent, however, in all the considered cases of the geodesic models or the string loop oscillation model of the twin HF QPOs, the black hole mass is clearly large enough to give a convincing argument that an intermediate black hole is located at the M82~X-1 source. The minimum of the predicted limits implies $M_{\rm M82 X-1}>130~M_{\odot}$. The minimum mass obtained for the models successful in the LMXB with stellar-mass black holes is $176~M_{\odot}$ implied by the RE model.
If the nodal model of the low frequency oscillations is simultaneously applied, the mass and spin of the black hole can be limited from above by relations $M_{\rm M82 X-1}<660~M_{\odot}$ and $a_{\rm M82 X-1}<0.6$. However, these limits are based on the assumption that some of the observed low frequency QPOs occurred at the radius where the simultaneous twin HF QPOs were generated in the M82~X-1 source. In the case when this assumption is incorrect, the upper limits could be shifted to large values. The string loop model places no constraint on the black hole spin, the M82~X-1 black hole mass range is calculated allowing any possible spin and reads $430~M_{\odot}<M_{\rm M82X-1}<860~M_{\odot}$. 

Note that discovery of simultaneity of one of the low frequency QPOs and the twin HF QPOs could significantly restrict the black hole mass and spin limits. On the other hand, additional data from measurements of spectral continuum or profiled spectral lines are necessary in order to distinct predictions of the various considered geodesic models of QPOs.  

\section*{Acknowledgments}

The authors would like to thank the Albert Einstein Centre for gravitation and astrophysics supported by the~Czech Science Foundation Grant No. 14-37086G. 


\bibliographystyle{mn2e}

\def\prc{Phys. Rev. C}
\def\pre{Phys. Rev. E}
\def\prd{Phys. Rev. D}
\def\jcap{Journal of Cosmology and Astroparticle Physics}
\def\apss{Ap\&SS}
\def\mnras{MNRAS}
\def\apj{AJ}
\def\aap{A\&A}
\def\actaa{Acta Astronomica}
\def\pasj{PASJ}
\def\apjl{AJ Letters}
\def\pasa{Publications Astronomical Society of Australia}
\def\nat{Nature}
\def\physrep{Phys. Rep.}
\def\araa{Annu. Rev. Astron. Astrophys.}
\def\apjs{ApJS}


\begin{thebibliography}{}

\bibitem[\protect\citeauthoryear{{Abramowicz}, {Jaroszynski} \&
  {Sikora}}{{Abramowicz} et~al.}{1978}]{Abr-etal:1978:ASTRA:}
{Abramowicz} M.,  {Jaroszynski} M.,    {Sikora} M.,  1978, \aap, 63, 221

\bibitem[\protect\citeauthoryear{{Abramowicz}, {Bulik}, {Bursa} \&
  {Klu{\'z}niak}}{{Abramowicz} et~al.}{2003}]{Abr-Bul-Bur-Klu:2003:ASTRA:}
{Abramowicz} M.~A.,  {Bulik} T.,  {Bursa} M.,    {Klu{\'z}niak} W.,  2003,
  \aap, 404, L21

\bibitem[\protect\citeauthoryear{{Abramowicz} \& {Klu{\'z}niak}}{{Abramowicz}
  \& {Klu{\'z}niak}}{2001}]{Abr-Klu:2001:AA:}
{Abramowicz} M.~A.,  {Klu{\'z}niak} W.,  2001, \aap, 374, L19

\bibitem[\protect\citeauthoryear{{Abramowicz}, {Klu{\'z}niak}, {McClintock} \&
  {Remillard}}{{Abramowicz} et~al.}{2004}]{Abr-etal:2004:ApJ:}
{Abramowicz} M.~A.,  {Klu{\'z}niak} W.,  {McClintock} J.~E.,    {Remillard}
  R.~A.,  2004, \apjl, 609, L63

\bibitem[\protect\citeauthoryear{{Aliev} \& {Galtsov}}{{Aliev} \&
  {Galtsov}}{1981}]{Ali-Gal:1981:GRG:}
{Aliev} A.~N.,  {Galtsov} D.~V.,  1981, General Relativity and Gravitation, 13,
  899

\bibitem[\protect\citeauthoryear{{Arnold}}{{Arnold}}{1978}]{Arnold:1978:book:}
{Arnold} V.~I.,  1978, {Mathematical methods of classical mechanics}.
New York: Springer

\bibitem[\protect\citeauthoryear{{Axelsson}, {Done} \&
  {Hjalmarsdotter}}{{Axelsson} et~al.}{2014}]{Axe-Don-Hja:2008:MNRAS:}
{Axelsson} M.,  {Done} C.,    {Hjalmarsdotter} L.,  2014, \mnras, 438, 657

\bibitem[\protect\citeauthoryear{{Bardeen}, {Press} \& {Teukolsky}}{{Bardeen}
  et~al.}{1972}]{Bar-Pre-Teu:1972:ApJ:}
{Bardeen} J.~M.,  {Press} W.~H.,    {Teukolsky} S.~A.,  1972, \apj, 178, 347

\bibitem[\protect\citeauthoryear{{Bursa}}{{Bursa}}{2005}]{Bur:2005:RAG:}
{Bursa} M.,  2005, in {Hled{\'{\i}}k} S.,  {Stuchl{\'{\i}}k} Z.,  eds, RAGtime
  6/7: Workshops on black holes and neutron stars {High-frequency QPOs in GRO
  J1655-40: Constraints on resonance models by spectral fits}.
pp 39--45

\bibitem[\protect\citeauthoryear{{Casella}, {Ponti}, {Patruno}, {Belloni},
  {Miniutti} \& {Zampieri}}{{Casella} et~al.}{2008}]{Cas-etal:2008:MNRAS:}
{Casella} P.,  {Ponti} G.,  {Patruno} A.,  {Belloni} T.,  {Miniutti} G.,
  {Zampieri} L.,  2008, \mnras, 387, 1707

\bibitem[\protect\citeauthoryear{{Christensson} \& {Hindmarsh}}{{Christensson}
  \& {Hindmarsh}}{1999}]{Chri-Hin:1999:PhRvD:}
{Christensson} M.,  {Hindmarsh} M.,  1999, \prd, 60, 063001

\bibitem[\protect\citeauthoryear{{Cremaschini} \&
  {Stuchl{\'{\i}}k}}{{Cremaschini} \&
  {Stuchl{\'{\i}}k}}{2013}]{Cre-Stu:2013:PhRvE:}
{Cremaschini} C.,  {Stuchl{\'{\i}}k} Z.,  2013, \pre, 87, 043113

\bibitem[\protect\citeauthoryear{{Cremaschini}, {Stuchl{\'{\i}}k} \&
  {Tessarotto}}{{Cremaschini} et~al.}{2013}]{Cre-Stu-Tes:2013:PlasmaPhys:}
{Cremaschini} C.,  {Stuchl{\'{\i}}k} Z.,    {Tessarotto} M.,  2013, Physics of
  Plasmas, 20, 052905

\bibitem[\protect\citeauthoryear{{Dewangan}, {Titarchuk} \&
  {Griffiths}}{{Dewangan} et~al.}{2006}]{Dew-Tit-Gri:2006:ApJ:}
{Dewangan} G.~C.,  {Titarchuk} L.,    {Griffiths} R.~E.,  2006, \apjl, 637, L21

\bibitem[\protect\citeauthoryear{{Feng} \& {Kaaret}}{{Feng} \&
  {Kaaret}}{2010}]{Fen-Kaa:2010:ApJ:}
{Feng} H.,  {Kaaret} P.,  2010, \apjl, 712, L169

\bibitem[\protect\citeauthoryear{{Feroci}, {den Herder}, {Bozzo}, {Barret},
  {Brandt}, {Hernanz}, {van der Klis}, {Pohl}, {Santangelo}, {Stella} \& et
  al.}{{Feroci} et~al.}{2012}]{Fer-etal:2012:ExpAstr:}
{Feroci} M.,  {den Herder} J.~W.,  {Bozzo} E.,  {Barret} D.,  {Brandt} S.,
  {Hernanz} M.,  {van der Klis} M.,  {Pohl} M.,  {Santangelo} A.,  {Stella} L.,
     et al. 2012, Experimental Astronomy, 34, 415

\bibitem[\protect\citeauthoryear{{Fu} \& {Lai}}{{Fu} \&
  {Lai}}{2009}]{Fu-Lai:2009:ApJ:}
{Fu} W.,  {Lai} D.,  2009, \apj, 690, 1386

\bibitem[\protect\citeauthoryear{{Fu} \& {Lai}}{{Fu} \&
  {Lai}}{2011}]{Fu-Lai:2011:MNRAS:}
{Fu} W.,  {Lai} D.,  2011, \mnras, 410, 399

\bibitem[\protect\citeauthoryear{{Hor{\'a}k}, {Abramowicz}, {Klu{\'z}niak},
  {Rebusco} \& {T{\"o}r{\"o}k}}{{Hor{\'a}k}
  et~al.}{2009}]{Hor-etal:2009:ASTRA:}
{Hor{\'a}k} J.,  {Abramowicz} M.~A.,  {Klu{\'z}niak} W.,  {Rebusco} P.,
  {T{\"o}r{\"o}k} G.,  2009, \aap, 499, 535

\bibitem[\protect\citeauthoryear{{Jacobson} \& {Sotiriou}}{{Jacobson} \&
  {Sotiriou}}{2009}]{Jac-Sot:2009:PHYSR4:}
{Jacobson} T.,  {Sotiriou} T.~P.,  2009, \prd, 79, 065029

\bibitem[\protect\citeauthoryear{{Kaaret}, {Prestwich}, {Zezas}, {Murray},
  {Kim}, {Kilgard}, {Schlegel} \& {Ward}}{{Kaaret}
  et~al.}{2001}]{Kaa-etal:2001:MNRAS:}
{Kaaret} P.,  {Prestwich} A.~H.,  {Zezas} A.,  {Murray} S.~S.,  {Kim} D.-W.,
  {Kilgard} R.~E.,  {Schlegel} E.~M.,    {Ward} M.~J.,  2001, \mnras, 321, L29

\bibitem[\protect\citeauthoryear{{Kato}}{{Kato}}{2004}]{Kat:2004:PASJ:}
{Kato} S.,  2004, \pasj, 56, 905

\bibitem[\protect\citeauthoryear{{Kato}}{{Kato}}{2008}]{Kat:2008:PASJ:}
{Kato} S.,  2008, \pasj, 60, 889

\bibitem[\protect\citeauthoryear{{Kato} \& {Fukue}}{{Kato} \&
  {Fukue}}{1980}]{Kat-Fuk:1980:PAJS:}
{Kato} S.,  {Fukue} J.,  1980, \pasj, 32, 377

\bibitem[\protect\citeauthoryear{{Kato}, {Fukue} \& {Mineshige}}{{Kato}
  et~al.}{1998}]{Kat-Fuk-Min:1998:BHAccDis:}
{Kato} S.,  {Fukue} J.,    {Mineshige} S.,  eds, 1998, {Black-hole accretion
  disks}

\bibitem[\protect\citeauthoryear{{Kolo{\v s}} \& {Stuchl{\'{\i}}k}}{{Kolo{\v
  s}} \& {Stuchl{\'{\i}}k}}{2010}]{Kol-Stu:2010:PHYSR4:}
{Kolo{\v s}} M.,  {Stuchl{\'{\i}}k} Z.,  2010, \prd, 82, 125012

\bibitem[\protect\citeauthoryear{{Kolo{\v s}} \& {Stuchl{\'{\i}}k}}{{Kolo{\v
  s}} \& {Stuchl{\'{\i}}k}}{2013}]{Kol-Stu:2013:PHYSR4:}
{Kolo{\v s}} M.,  {Stuchl{\'{\i}}k} Z.,  2013, \prd, 88, 065004

\bibitem[\protect\citeauthoryear{{Kosti{\'c}}, {{\v C}ade{\v z}}, {Calvani} \&
  {Gomboc}}{{Kosti{\'c}} et~al.}{2009}]{Kos-etal:2009:ASTRA:}
{Kosti{\'c}} U.,  {{\v C}ade{\v z}} A.,  {Calvani} M.,    {Gomboc} A.,  2009,
  \aap, 496, 307

\bibitem[\protect\citeauthoryear{{Kov{\'a}{\v r}}}{{Kov{\'a}{\v
  r}}}{2013}]{Kov:2013:EPJP:}
{Kov{\'a}{\v r}} J.,  2013, European Physical Journal Plus, 128, 142

\bibitem[\protect\citeauthoryear{{Kozlowski}, {Jaroszynski} \&
  {Abramowicz}}{{Kozlowski} et~al.}{1978}]{Koz-etal:1977:ASTRA:}
{Kozlowski} M.,  {Jaroszynski} M.,    {Abramowicz} M.~A.,  1978, \aap, 63, 209

\bibitem[\protect\citeauthoryear{{Landau} \& {Lifshitz}}{{Landau} \&
  {Lifshitz}}{1969}]{Lan-Lif:1969:Mech:}
{Landau} L.~D.,  {Lifshitz} E.~M.,  1969, {Mechanics}.
Oxford: Pergamon Press

\bibitem[\protect\citeauthoryear{{Larsen}}{{Larsen}}{1993}]{Lar:1993:CLAQG:}
{Larsen} A.~L.,  1993, Classical and Quantum Gravity, 10, 1541

\bibitem[\protect\citeauthoryear{{Larsen}}{{Larsen}}{1994}]{Lar:1994:CLAQG:}
{Larsen} A.~L.,  1994, Classical and Quantum Gravity, 11, 1201

\bibitem[\protect\citeauthoryear{{Machida} \& {Matsumoto}}{{Machida} \&
  {Matsumoto}}{2008}]{Mac-Mat:2008:PAJC:}
{Machida} M.,  {Matsumoto} R.,  2008, \pasj, 60, 613

\bibitem[\protect\citeauthoryear{{Matsumoto}, {Tsuru}, {Koyama}, {Awaki},
  {Canizares}, {Kawai}, {Matsushita} \& {Kawabe}}{{Matsumoto}
  et~al.}{2001}]{Mat-etal:2001:ApJ:}
{Matsumoto} H.,  {Tsuru} T.~G.,  {Koyama} K.,  {Awaki} H.,  {Canizares} C.~R.,
  {Kawai} N.,  {Matsushita} S.,    {Kawabe} R.,  2001, \apjl, 547, L25

\bibitem[\protect\citeauthoryear{{Misner}, {Thorne} \& {Wheeler}}{{Misner}
  et~al.}{1973}]{Mis-Tho-Whe:1973:Gra:}
{Misner} C.~W.,  {Thorne} K.~S.,    {Wheeler} J.~A.,  1973, {Gravitation}

\bibitem[\protect\citeauthoryear{{Montero} \& {Zanotti}}{{Montero} \&
  {Zanotti}}{2012}]{Mon-Zan:2012:MNRAS:}
{Montero} P.~J.,  {Zanotti} O.,  2012, \mnras, 419, 1507

\bibitem[\protect\citeauthoryear{{Motta}, {Mu{\~n}oz-Darias}, {Sanna},
  {Fender}, {Belloni} \& {Stella}}{{Motta} et~al.}{2014}]{Mot-etal:2014:MNRAS:}
{Motta} S.~E.,  {Mu{\~n}oz-Darias} T.,  {Sanna} A.,  {Fender} R.,  {Belloni}
  T.,    {Stella} L.,  2014, \mnras, 439, L65

\bibitem[\protect\citeauthoryear{{Mucciarelli}, {Casella}, {Belloni},
  {Zampieri} \& {Ranalli}}{{Mucciarelli} et~al.}{2006}]{Muc-etal:2006:MNRAS:}
{Mucciarelli} P.,  {Casella} P.,  {Belloni} T.,  {Zampieri} L.,    {Ranalli}
  P.,  2006, \mnras, 365, 1123

\bibitem[\protect\citeauthoryear{{Nayfeh} \& {Mook}}{{Nayfeh} \&
  {Mook}}{1979}]{Nay-Moo:1979:NonOscilations:}
{Nayfeh} A.~H.,  {Mook} D.~T.,  1979, {Nonlinear oscillations}.
New York : Wiley

\bibitem[\protect\citeauthoryear{{Novikov} \& {Thorne}}{{Novikov} \&
  {Thorne}}{1973}]{Nov-Tho:1973:BlaHol:}
{Novikov} I.~D.,  {Thorne} K.~S.,  1973, in {Dewitt} C.,  {Dewitt} B.~S.,  eds,
  Black Holes (Les Astres Occlus) {Astrophysics of black holes.}.
pp 343--450

\bibitem[\protect\citeauthoryear{{Nowak} \& {Lehr}}{{Nowak} \&
  {Lehr}}{1998}]{Now-Leh:1998:TBHAD:}
{Nowak} M.~A.,  {Lehr} D.~E.,  1998, in {Abramowicz} M.~A.,  {Bj{\"o}rnsson}
  G.,   {Pringle} J.~E.,  eds, Theory of Black Hole Accretion Disks {Stable
  oscillations of black hole accretion discs.}.
pp 233--253

\bibitem[\protect\citeauthoryear{{Okajima}, {Ebisawa} \& {Kawaguchi}}{{Okajima}
  et~al.}{2006}]{Oka-Epi-Kaw:2006:ApJ:}
{Okajima} T.,  {Ebisawa} K.,    {Kawaguchi} T.,  2006, \apjl, 652, L105

\bibitem[\protect\citeauthoryear{{Page} \& {Thorne}}{{Page} \&
  {Thorne}}{1974}]{Pag-Tho:1974:ApJ:}
{Page} D.~N.,  {Thorne} K.~S.,  1974, \apj, 191, 499

\bibitem[\protect\citeauthoryear{{Pasham} \& {Strohmayer}}{{Pasham} \&
  {Strohmayer}}{2013}]{Pas-Str:2013:ApJ:}
{Pasham} D.~R.,  {Strohmayer} T.~E.,  2013, \apj, 771, 101

\bibitem[\protect\citeauthoryear{{Pasham}, {Strohmayer} \&
  {Mushotzky}}{{Pasham} et~al.}{2014}]{Pas-Str-Mus:2015:Nat:}
{Pasham} D.~R.,  {Strohmayer} T.~E.,    {Mushotzky} R.~F.,  2014, \nat, 513, 74

\bibitem[\protect\citeauthoryear{{Pawar}, {Kalamkar}, {Altamirano}, {Linares},
  {Shanthi}, {Strohmayer}, {Bhattacharya} \& {van der Klis}}{{Pawar}
  et~al.}{2013}]{Paw-etal:2013:MONRAS:}
{Pawar} D.~D.,  {Kalamkar} M.,  {Altamirano} D.,  {Linares} M.,  {Shanthi} K.,
  {Strohmayer} T.,  {Bhattacharya} D.,    {van der Klis} M.,  2013, \mnras,
  433, 2436

\bibitem[\protect\citeauthoryear{{Reynolds} \& {Miller}}{{Reynolds} \&
  {Miller}}{2009}]{Rey-Col:2009:ApJ:}
{Reynolds} C.~S.,  {Miller} M.~C.,  2009, \apj, 692, 869

\bibitem[\protect\citeauthoryear{{Rezzolla}, {Yoshida}, {Maccarone} \&
  {Zanotti}}{{Rezzolla} et~al.}{2003}]{Rez-etal:2003:MNRAS:}
{Rezzolla} L.,  {Yoshida} S.,  {Maccarone} T.~J.,    {Zanotti} O.,  2003,
  \mnras, 344, L37

\bibitem[\protect\citeauthoryear{{Semenov}, {Dyadechkin} \& {Punsly}}{{Semenov}
  et~al.}{2004}]{Sem-Dya-Pun:2004:Sci:}
{Semenov} V.,  {Dyadechkin} S.,    {Punsly} B.,  2004, Science, 305, 978

\bibitem[\protect\citeauthoryear{{Semenov} \& {Bernikov}}{{Semenov} \&
  {Bernikov}}{1991}]{Sem-Ber:1990:ASS:}
{Semenov} V.~S.,  {Bernikov} L.~V.,  1991, \apss, 184, 157

\bibitem[\protect\citeauthoryear{{Stella} \& {Vietri}}{{Stella} \&
  {Vietri}}{1998}]{Ste-Vie:1998:ApJ:}
{Stella} L.,  {Vietri} M.,  1998, \apjl, 492, L59

\bibitem[\protect\citeauthoryear{{Stella} \& {Vietri}}{{Stella} \&
  {Vietri}}{1999}]{Ste-Vie:1999:PHYSRL:}
{Stella} L.,  {Vietri} M.,  1999, Physical Review Letters, 82, 17

\bibitem[\protect\citeauthoryear{{Stella}, {Vietri} \& {Morsink}}{{Stella}
  et~al.}{1999}]{Ste-Vie-Mor:1999:ApJ:}
{Stella} L.,  {Vietri} M.,    {Morsink} S.~M.,  1999, \apjl, 524, L63

\bibitem[\protect\citeauthoryear{{Straub} \& {{\v S}r{\'a}mkov{\'a}}}{{Straub}
  \& {{\v S}r{\'a}mkov{\'a}}}{2009}]{Str-Sra:2009:CLAQG:}
{Straub} O.,  {{\v S}r{\'a}mkov{\'a}} E.,  2009, Classical and Quantum Gravity,
  26, 055011

\bibitem[\protect\citeauthoryear{{Strohmayer} \& {Mushotzky}}{{Strohmayer} \&
  {Mushotzky}}{2003}]{Str-Mus:2003:ApJ:}
{Strohmayer} T.~E.,  {Mushotzky} R.~F.,  2003, \apjl, 586, L61

\bibitem[\protect\citeauthoryear{Stuchl{\'{\i}}k}{Stuchl{\'{\i}}k}{1983}]{Stu:1983:BULAI:}
Stuchl{\'{\i}}k Z.,  1983, Bulletin of the Astronomical Institutes of
  Czechoslovakia, 34, 129

\bibitem[\protect\citeauthoryear{{Stuchl{\'{\i}}k} \&
  {Hled{\'{\i}}k}}{{Stuchl{\'{\i}}k} \&
  {Hled{\'{\i}}k}}{1999}]{Stu-Hle:1999:PHYSR4:}
{Stuchl{\'{\i}}k} Z.,  {Hled{\'{\i}}k} S.,  1999, \prd, 60, 044006

\bibitem[\protect\citeauthoryear{{Stuchl{\'{\i}}k} \& {Kolo{\v
  s}}}{{Stuchl{\'{\i}}k} \& {Kolo{\v s}}}{2012a}]{Stu-Kol:2012:PHYSR4:}
{Stuchl{\'{\i}}k} Z.,  {Kolo{\v s}} M.,  2012a, \prd, 85, 065022

\bibitem[\protect\citeauthoryear{{Stuchl{\'{\i}}k} \& {Kolo{\v
  s}}}{{Stuchl{\'{\i}}k} \& {Kolo{\v s}}}{2012b}]{Stu-Kol:2012:JCAP:}
{Stuchl{\'{\i}}k} Z.,  {Kolo{\v s}} M.,  2012b, \jcap, 10, 8

\bibitem[\protect\citeauthoryear{{Stuchl{\'{\i}}k} \& {Kolo{\v
  s}}}{{Stuchl{\'{\i}}k} \& {Kolo{\v s}}}{2014}]{Stu-Kol:2014:PHYSR4:}
{Stuchl{\'{\i}}k} Z.,  {Kolo{\v s}} M.,  2014, \prd, 89, 065007

\bibitem[\protect\citeauthoryear{{Stuchl{\'{\i}}k} \& {Kolo{\v
  s}}}{{Stuchl{\'{\i}}k} \& {Kolo{\v s}}}{2015}]{Stu-Kol:2015:GRG:}
{Stuchl{\'{\i}}k} Z.,  {Kolo{\v s}} M.,  2015, General Relativity and
  Gravitation, 47, 1

\bibitem[\protect\citeauthoryear{{Stuchl{\'{\i}}k}, {Konar}, {Miller} \&
  {Hled{\'{\i}}k}}{{Stuchl{\'{\i}}k}
  et~al.}{2008}]{Stu-Kon-Mil-Hle:2008:ASTRA:}
{Stuchl{\'{\i}}k} Z.,  {Konar} S.,  {Miller} J.~C.,    {Hled{\'{\i}}k} S.,
  2008, \aap, 489, 963

\bibitem[\protect\citeauthoryear{{Stuchl{\'{\i}}k}, {Kotrlov{\'a}} \&
  {T{\"o}r{\"o}k}}{{Stuchl{\'{\i}}k} et~al.}{2011}]{Stu-Kot-Tor:2011:ASTRA:}
{Stuchl{\'{\i}}k} Z.,  {Kotrlov{\'a}} A.,    {T{\"o}r{\"o}k} G.,  2011, \aap,
  525, A82

\bibitem[\protect\citeauthoryear{{Stuchl{\'{\i}}k}, {Kotrlov{\'a}} \&
  {T{\"o}r{\"o}k}}{{Stuchl{\'{\i}}k} et~al.}{2012}]{Stu-Kot-Tor:2012:ACTA:}
{Stuchl{\'{\i}}k} Z.,  {Kotrlov{\'a}} A.,    {T{\"o}r{\"o}k} G.,  2012, \actaa,
  62, 389

\bibitem[\protect\citeauthoryear{{Stuchl{\'{\i}}k}, {Kotrlov{\'a}} \&
  {T{\"o}r{\"o}k}}{{Stuchl{\'{\i}}k} et~al.}{2013}]{Stu-Kot-Tor:2013:ASTRA:}
{Stuchl{\'{\i}}k} Z.,  {Kotrlov{\'a}} A.,    {T{\"o}r{\"o}k} G.,  2013, \aap,
  552, A10

\bibitem[\protect\citeauthoryear{{Stuchl{\'{\i}}k} \&
  {Schee}}{{Stuchl{\'{\i}}k} \& {Schee}}{2012}]{Stu-Sche:2012:CLAQG:}
{Stuchl{\'{\i}}k} Z.,  {Schee} J.,  2012, Classical and Quantum Gravity, 29,
  065002

\bibitem[\protect\citeauthoryear{{Stuchl{\'{\i}}k}, {Slan{\'y}} \& {Kov{\'a}{\v
  r}}}{{Stuchl{\'{\i}}k} et~al.}{2009}]{Stu-etal:2009:CLAQG:}
{Stuchl{\'{\i}}k} Z.,  {Slan{\'y}} P.,    {Kov{\'a}{\v r}} J.,  2009, Classical
  and Quantum Gravity, 26, 215013

\bibitem[\protect\citeauthoryear{{Stuchl{\'{\i}}k}, {Slan{\'y}} \&
  {T{\"o}r{\"o}k}}{{Stuchl{\'{\i}}k} et~al.}{2007}]{Stu-Sla-Tor:2007:ASTRA:}
{Stuchl{\'{\i}}k} Z.,  {Slan{\'y}} P.,    {T{\"o}r{\"o}k} G.,  2007, \aap, 463,
  807

\bibitem[\protect\citeauthoryear{{Stuchl{\'{\i}}k}, {Slan{\'y}},
  {T{\"o}r{\"o}k} \& {Abramowicz}}{{Stuchl{\'{\i}}k}
  et~al.}{2005}]{Stu-etal:2005:PHYSR4:}
{Stuchl{\'{\i}}k} Z.,  {Slan{\'y}} P.,  {T{\"o}r{\"o}k} G.,    {Abramowicz}
  M.~A.,  2005, \prd, 71, 024037

\bibitem[\protect\citeauthoryear{{T{\"o}r{\"o}k}, {Abramowicz}, {Klu{\'z}niak}
  \& {Stuchl{\'{\i}}k}}{{T{\"o}r{\"o}k} et~al.}{2005}]{Tor-etal:2005:ASTRA:}
{T{\"o}r{\"o}k} G.,  {Abramowicz} M.~A.,  {Klu{\'z}niak} W.,
  {Stuchl{\'{\i}}k} Z.,  2005, \aap, 436, 1

\bibitem[\protect\citeauthoryear{{T{\"o}r{\"o}k}, {Kotrlov{\'a}}, {{\v
  S}r{\'a}mkov{\'a}} \& {Stuchl{\'{\i}}k}}{{T{\"o}r{\"o}k}
  et~al.}{2011}]{Tor-etal:2011:ASTRA:}
{T{\"o}r{\"o}k} G.,  {Kotrlov{\'a}} A.,  {{\v S}r{\'a}mkov{\'a}} E.,
  {Stuchl{\'{\i}}k} Z.,  2011, \aap, 531, A59

\bibitem[\protect\citeauthoryear{{T{\"o}r{\"o}k} \&
  {Stuchl{\'{\i}}k}}{{T{\"o}r{\"o}k} \&
  {Stuchl{\'{\i}}k}}{2005}]{Tor-Stu:2005:ASTRA:}
{T{\"o}r{\"o}k} G.,  {Stuchl{\'{\i}}k} Z.,  2005, \aap, 437, 775

\bibitem[\protect\citeauthoryear{{Tursunov}, {Kolo{\v s}}, {Ahmedov} \&
  {Stuchl{\'{\i}}k}}{{Tursunov} et~al.}{2013}]{Tur-etal:2013:PHYSR4:}
{Tursunov} A.,  {Kolo{\v s}} M.,  {Ahmedov} B.,    {Stuchl{\'{\i}}k} Z.,  2013,
  \prd, 87, 125003

\bibitem[\protect\citeauthoryear{{Tursunov}, {Kolo{\v s}}, {Stuchl{\'{\i}}k} \&
  {Ahmedov}}{{Tursunov} et~al.}{2014}]{Tur-etal:2014:PHYSR4:}
{Tursunov} A.,  {Kolo{\v s}} M.,  {Stuchl{\'{\i}}k} Z.,    {Ahmedov} B.,  2014,
  \prd, 90, 085009

\bibitem[\protect\citeauthoryear{{{\v C}ade{\v z}}, {Calvani} \&
  {Kosti{\'c}}}{{{\v C}ade{\v z}} et~al.}{2008}]{Cad-Cal-Kos:2008:ASTRA:}
{{\v C}ade{\v z}} A.,  {Calvani} M.,    {Kosti{\'c}} U.,  2008, \aap, 487, 527

\bibitem[\protect\citeauthoryear{{van der Klis}}{{van der
  Klis}}{2006}]{Kli:2006:Compact:}
{van der Klis} M.,  2006, {Rapid X-ray Variability}.
Cambridge University Press, pp 39--112

\bibitem[\protect\citeauthoryear{{Wagoner}}{{Wagoner}}{1999}]{Wag:1999:PHYSR:}
{Wagoner} R.~V.,  1999, \physrep, 311, 259

\bibitem[\protect\citeauthoryear{{Wagoner}, {Silbergleit} \&
  {Ortega-Rodr{\'{\i}}guez}}{{Wagoner} et~al.}{2001}]{Wag-etal:2001:ApJ:}
{Wagoner} R.~V.,  {Silbergleit} A.~S.,    {Ortega-Rodr{\'{\i}}guez} M.,  2001,
  \apjl, 559, L25

\bibitem[\protect\citeauthoryear{{Zanotti}, {Font}, {Rezzolla} \&
  {Montero}}{{Zanotti} et~al.}{2005}]{Zan-etal:2005:MNRAS:}
{Zanotti} O.,  {Font} J.~A.,  {Rezzolla} L.,    {Montero} P.~J.,  2005, \mnras,
  356, 1371

\end{thebibliography}

\end{document}